\documentclass[11pt,a4paper]{article}
\pdfoutput=1
\usepackage{cancel}
\usepackage{enumitem}
\usepackage{etex}
\usepackage{amsthm}
\usepackage{geometry}
\usepackage{dcolumn}
\usepackage{epsf}
\usepackage{mathrsfs}
\usepackage{multirow}
\usepackage{booktabs}
\usepackage{tabularx}
\usepackage{array}
\usepackage{slashed}
\usepackage{float}
\usepackage{leftidx}
\usepackage{setspace}
\usepackage{verbatim}
\usepackage{adjustbox}
\usepackage{tikz}
\usepackage[caption=false]{subfig}
\usepackage{arydshln}
\usepackage{jheppub}
\usepackage{shuffle}
\usepackage{amsmath}
\usepackage{cases}

\numberwithin{equation}{section}
\usetikzlibrary{arrows,decorations.markings,shapes.arrows,patterns,positioning}

\newcommand{\bea}{\begin{eqnarray}}
\newcommand{\eea}{\end{eqnarray}}
\newcommand{\bean}{\begin{eqnarray*}}
\newcommand{\eean}{\end{eqnarray*}}
\newcommand{\nn}{\nonumber\\}
\newcommand{\Sl}{\sum\limits}

\def\Label#1{\label{#1}%
  \smash{\hbox to0pt{\raise1ex\hbox{\tiny[#1]}\hss}}}
\def\Label#1{\label{#1}}
\renewcommand{\eqref}[1]{eq.~(\ref{#1})}
\newcommand{\figref}[1]{Fig.~\ref{#1}}

\newcommand{\secref}[1]{section~\ref{#1}}
\newcommand{\appref}[1]{appendix~\ref{#1}}

\def\Sl{\sum\limits}

\newcommand{\ctobedelete}[1]{}
\allowdisplaybreaks

\title{A graphic approach to identities induced from multi-trace Einstein-Yang-Mills amplitudes}

\author[a,b,c]{Yi-Jian Du} \author[a]{Linghui Hou}

\affiliation[a]{Center for Theoretical Physics, School of Physics and Technology,
Wuhan University, \\
No.299 Bayi Road, Wuhan 430072, China}
\affiliation[b]{College of Science, Tibet University,\\ No.10 Zangda East Road, Lasa, 850000, China}
\affiliation[c]{Suzhou Institute of Wuhan University,\\
No.377 Linquan Street, Suzhou, 215123, China}

\emailAdd{yijian.du@whu.edu.cn,hlh@whu.edu.cn}

\date{\today}
\abstract{Symmetries of Einstein-Yang-Mills (EYM) amplitudes, together with the recursive expansions, induce nontrivial identities for pure Yang-Mills amplitudes. In the previous work \cite{Hou:2018bwm}, we have already proven that the identities induced from tree level single-trace EYM amplitudes can be precisely expanded in terms of BCJ relations. In this paper, we extend the discussions to those identities induced from all tree level \emph{multi-trace} EYM amplitudes. Particularly, we establish a refined graphic rule for multi-trace EYM amplitudes and then show that the induced identities can be fully decomposed in terms of BCJ relations.
}

\begin{document}
\maketitle \flushbottom

\section{Introduction}
It has been proven that any tree level multi-trace Einstein-Yang-Mills (EYM) amplitude $A^{(m,s)}$ with $m$ gluon traces and $s$ gravitons can be recursively expanded in terms of the amplitudes $A^{(m',s')}$ with $s'+m'<s+m$ \cite{Du:2017gnh}. In the special case $m=1$, these relations give rise to the earlier proposed expansions of single-trace EYM amplitudes \cite{Fu:2017uzt,Chiodaroli:2017ngp,Teng:2017tbo}. When the recursive expansions are applied repeatedly until there is no graviton and only one gluon trace,  an arbitrary tree-level EYM amplitude is finally expressed as a combination of tree level color-ordered Yang-Mills (YM) ones. Such pure-YM expansions precisely coincide with the earlier studies on amplitudes with only a few gravitons and/or gluon traces \cite{Stieberger:2016lng,Nandan:2016pya,delaCruz:2016gnm,Schlotterer:2016cxa}.


The pure YM expansions of EYM amplitudes, together with the gauge invariance conditions of gravitons or the cyclic symmetries of gluon traces, induce nontrivial identities for color-ordered YM amplitudes \cite{Du:2017gnh}. These identities guaranteed the localities in the Britto-Cachazo-Feng-Witten (BCFW) \cite{Britto:2004ap,Britto:2005fq} proof of the recursive expansions \cite{Fu:2017uzt,Du:2017gnh} and played a crucial role in the proof \cite{Du:2018khm} of the equivalence between distinct approaches \cite{Du:2016tbc,Carrasco:2016ldy,Du:2017kpo} to nonlinear sigma model amplitudes.

A prominent feature of the identities induced from multi-trace EYM amplitudes is that the coefficients therein generally contain factors of all the three types of Lorentz contractions  $\epsilon\cdot\epsilon$, $\epsilon\cdot k$ and $k\cdot k$ where $\epsilon^{\mu}$ are half polarizations of gravitons and $k^{\mu}$ are external momenta. This is quite different from the known Kleiss-Kuijf (KK) \cite{Kleiss:1988ne} and Bern-Carrasco-Johansson (BCJ) relations \cite{Bern:2008qj} whose coefficients at most involve $k\cdot k$ factors. Nevertheless, several clues imply that the identities induced from multi-trace EYM amplitudes can be related with BCJ relations: (i). First, the relationship between the identities induced from single-trace EYM amplitudes and BCJ relations have already been founded \cite{Hou:2018bwm}, while the multi-trace amplitudes can be obtained through replacing gravitons by gluon traces in an appropriate way (as pointed in \cite{Du:2017gnh}). (ii). Second, as demonstrated in \cite{Du:2017gnh}, examples with a few gluon traces and gravitons provided evidence of the connection between the induced identities and BCJ relations. (iii). Third, the fact that color-kinematic duality \cite{Bern:2008qj}, which is the underlaid structure of the BCJ relations for YM amplitudes, can be resulted from gauge invariance \cite{Arkani-Hamed:2016rak} also implies that the identities induced from gauge invariance can be related with BCJ relations.


In the current paper,  we extend our discussions in \cite{Hou:2018bwm} to multi-trace cases and show that all these identities induced from multi-trace EYM amplitudes can be expanded in terms of BCJ relations. The main idea is sketched as follows:
\begin{itemize}
\item [\bf{(i).}] {\bf Refined graphic rule~} We first introduce a refined graphic rule\footnote{Graphic approaches to scattering amplitudes can be found in many literatures including: graphic study on Cachazo-He-Yuan \cite{Cachazo:2013gna,Cachazo:2013hca,Cachazo:2013iea,Cachazo:2014xea} formulas (see e.g. \cite{Lam:2015sqb,Lam:2016tlk,Bjerrum-Bohr:2016juj,Huang:2017ydz,Gao:2017dek,He:2018pue,Lam:2018tgm,Lam:2019mfk}), graphic construction  for BCJ numerators \cite{Du:2017kpo} and the refined graphic rule for single-trace EYM amplitudes \cite{Hou:2018bwm}.} for the coefficients in pure-YM expansions of multi-trace EYM amplitudes,  where the three types of factors $\epsilon\cdot\epsilon$, $\epsilon\cdot k$ and $k\cdot k$ are presented by distinct types of lines (as already introduced in the study of single-trace case \cite{Hou:2018bwm}) and a new type of line is invented to record the relative order of gluons in a gluon trace. The induced identities are then expressed through a summation over connected tree graphs which are built of the four types of lines, while color-ordered YM amplitudes corresponding to each graph can be collected in a proper way.

\item [\bf{(ii).}] {\bf Skeletons and components~} To relate the induced identities with BCJ relations, we split gluon traces in an appropriate way and then remove all the $k\cdot k$ lines from the graphs. After that, a \emph{physical graph} (i.e. a graph defined by the refined graphic rule) turns to a disconnected one which is called \emph{skeleton} and consists of disjoint\emph{ components}. The summation over all physical graphs is thus given by summing over all skeletons and summing over all the physical graphs corresponding to a given skeleton.

\item [\bf{(iii).}] {\bf The final upper and lower blocks~} For a given skeleton, the summation over all possible physical graphs can further be arranged by the following two steps: (a). Connect components via $k\cdot k$ lines properly such that the skeleton becomes a graph with only two disjoint maximally connected subgraphs, which are called \emph{the final upper and lower blocks},  (b). Connect the final upper and lower blocks into a physical graph via a $k\cdot k$ line. \emph{Spurious graphs}, which are not defined by the refined graphic rule, can also be introduced for a given configuration of the final upper and lower blocks. When associated with proper signs, all spurious graphs cancel out. Then the summation over all physical graphs for a given final upper and lower blocks can be reexpressed by a summation over all physical and spurious graphs.

\item [\bf{(iv).}] {\bf Expressing an induced identity by BCJ relations~} We finally find that all contributions of the  physical and spurious graphs, corresponding to a given configuration of the final upper and lower blocks, together can be written as a combination of the graph-based BCJ relations \cite{Hou:2018bwm}  which have been proven to be combinations of the traditional BCJ relations \cite{Bern:2008qj}\footnote{In this paper, we distinguish the earlier proposed forms of BCJ relations (see \cite{Bern:2008qj,BjerrumBohr:2009rd, Chen:2011jxa}) from the graph-based BCJ relations \cite{Hou:2018bwm} by the name \emph{traditional BCJ relations}.} .

\end{itemize}


The structure of this paper is the following. In \secref{Sec:RefinedGraphicRule}, we introduce a refined graphic rule for the expansion of multi-trace EYM amplitudes and then express the induced identities by this rule. We further show two examples in \secref{Sec:Examples} which support the fact that the contributions of all those graphs corresponding to a given skeleton can be written as a combination of graph-based BCJ relations. In \secref{sec:GeneralStudy}, the general pattern of skeletons and components are studied. We then provide the general construction of the final upper and lower blocks for a given skeleton. The pattern of spurious graphs is also discussed. We finally show how to express the contributions of all (physical and spurious) graphs for a given final upper and lower blocks in terms of the graph-based BCJ relations. This work is summarized in \secref{sec:Conclusions}. A review of the background knowledge, the proof of the splitting trace relation and the pattern of the signs for graphs are included in the appendix.


\subsubsection*{Convention of notations}
The notations in this paper are gathered as follows.

{\bf Permutations and sets}: Permutations are denoted by boldface Greek letters: $\pmb{\sigma}$, $\pmb{\alpha}$, $\pmb{\beta}$, $\pmb{\gamma}$, $\pmb{\zeta}$, etc. The $i$-th element in $\pmb{\sigma}$ is denoted by $\sigma(i)$. The position of an element $a$ in $\pmb{\sigma}$ is expressed by $\sigma^{-1}(a)$. The inverse permutation of elements in an ordered set $\pmb{X}$ is denoted by $\pmb{X}^{T}$.
Shuffle permutations of two ordered sets $\pmb{X}$ and $\pmb{Y}$ are written as $\pmb{X}\shuffle\pmb{Y}$. The number of elements in set $\pmb{A}$ is presented by $|\pmb{A}|$. We use $\pmb{A}\setminus\pmb{B}$ to denote the difference of the sets $\pmb{A}$ and $\pmb{B}$.

{\bf Gravitons and gluon traces:} Gluon traces are denoted by boldface numbers $\pmb{1}$, $\pmb{2}$, ...,  or boldface lowercase Latin letters $\pmb{t}$, $\pmb{i}$ ... If a trace $\pmb{t}_i$ can be written as $\pmb{t}_i=\{a_i,\pmb{X}_i,b_i,\pmb{Y}_i\}$ ( $\pmb{X}_i$ and $\pmb{Y}_i$ are the two ordered sets of gluons which are separated by the gluons $a_i$ and $b_i$), we define $\mathsf{KK}[\pmb{t}_i,a_i,b_i]\equiv\pmb{X}_i\shuffle \pmb{Y}_i^T$ and $(-1)^{|\pmb{t}_i,a_i,b_i|}\equiv(-1)^{|\pmb{Y}_i|}$.
The notation $\mathsf{H}$ ( $\mathsf{H}=\{h_1,h_2,\dots, h_s\}$) stands for a graviton set with gravitons $h_1$, ..., $h_s$. The set of gravitons $h_1, \dots, h_s$ and gluon traces $\pmb{2},\dots,\pmb{m}$ is denoted by $\pmb{\mathcal{H}}$ ($\pmb{\mathcal{H}}= \{\pmb{2},\dots,\pmb{m}, h_1, \dots, h_s\}$). The multi-trace EYM amplitude with traces $\pmb{1}=\{1,\dots,r\},\pmb{2},\dots,\pmb{m}$ and graviton set $\mathsf{H}$ is written as $A(1,2,\ldots, r|\pmb{2}|\ldots|\pmb{m}\Vert\mathsf{H})$.

{\textbf{Graphs:}} Graphs are denoted by $\mathcal{F}$, $\mathcal{G}$ or $\mathcal{T}$. The notation $\mathcal{G}'$ stands for the skeleton of a graph $\mathcal{G}$. Reference order and root set are respectively expressed by $\mathsf{R}$ and $\mathcal{R}$. Components of a skeleton are given by $\mathscr{A}$, $\mathscr{B}$, $\mathscr{C}$ ..., while a chain of components is denoted by $\mathbb{CH}$. The reference order of components is given by $\mathsf{R}_{\mathscr{C}}$. The final upper and lower blocks are respectively presented by $\mathscr{U}$ and $\mathscr{L}$ whose disjoint union is $\mathscr{U}\oplus\mathscr{L}$.

\section{Refined graphic rule for multi-trace EYM amplitudes and the induced identities}\label{Sec:RefinedGraphicRule}
In this section, we present a \emph{refined graphic rule}, by which one expresses a tree level multi-trace EYM amplitude $A(1,2,\ldots, r|\pmb{2}|\ldots|\pmb{m}\Vert\mathsf{H})$ with $m$ gluon traces $\pmb{1}\equiv\{1,2,\ldots,r\},\pmb{2},\ldots,\pmb{m}$ and $s$ gravitons $\mathsf{H}\equiv\{h_1,h_2,\dots, h_s\}$ in terms of $(m+s)$-point tree level color-ordered YM amplitudes:
\bea
\boxed{A(1,2,\ldots, r|\pmb{2}|\ldots|\pmb{m}\Vert\mathsf{H})=\Sl_{\mathcal{F}} \mathcal{C}^{\mathcal{F}}\left[\Sl_{\pmb{\sigma}^{\mathcal{F}}}A(1,\pmb{\sigma}^{\mathcal{F}},r)\right]},\Label{Eq:PureYMExpansion}
\eea
where we have summed over all possible connected tree graphs $\mathcal{F}$. Each graph $\mathcal{F}$ defines a coefficient $\mathcal{C}^{\mathcal{F}}$ and proper permutations $1,\pmb{\sigma}^{\mathcal{F}},r$ (of all elements in $\pmb{1}\cup\pmb{2}\cup\ldots\cup\pmb{m}\cup\mathsf{H}$) according to the refined graphic rule.
The expansion (\ref{Eq:PureYMExpansion}) is obtained by applying the recursive expansion (\ref{Eq:MultiTrace}) (see \cite{Du:2017gnh}) iteratively and it is essentially equivalent to the graphic expansion given in \cite{Du:2017gnh}. Two examples which are helpful for understanding the refined graphic rule are given in this section. We then provide two identities that are respectively induced by the gauge invariance condition of a graviton and the cyclic symmetry of a gluon trace.

\subsection{Refined graphic rule}

To illustrate the refined graphic rule for the expansion (\ref{Eq:PureYMExpansion}), it is helpful to consider the gluon trace $\pmb{1}$ in the EYM amplitude $A(1,2,\ldots, r|\pmb{2}|\ldots|\pmb{m}\Vert\mathsf{H})$ as a special one and denote the set of other gluon traces $\pmb{2},\ldots,\pmb{m}$ and all gravitons $h_1, \dots, h_s$ by  $\pmb{\mathcal{H}}\equiv \{\pmb{2},\dots,\pmb{m}, h_1, \dots, h_s\}$. In the set $\pmb{\mathcal{H}}$, each gluon trace is always considered as a single object. The refined graphic rule is then expressed as follows:

{\noindent\bf{~Step-1}} Define a reference order of elements in $\pmb{\mathcal{H}}$ as the following ordered set:
\bea
\mathsf{R}=\left\{\mathcal{H}_{\rho(1)},\mathcal{H}_{\rho(2)},\dots,\mathcal{H}_{\rho(l=m+s-1)}\right\},\Label{Eq:ReferenceOrder}
\eea
where each $\mathcal{H}_{i}$ stands for an element (i.e. a graviton or a gluon trace) of $\pmb{\mathcal{H}}$ and the position of $\mathcal{H}_{i}$ in $\mathsf{R}$ is called its \emph{weight}. Apparently, $\mathcal{H}_{\rho(l=m+s-1)}$ is the highest-weight element in the reference order (\ref{Eq:ReferenceOrder}), while $\mathcal{H}_{\rho(1)}$ is the lowest-weight one. We also define the root set $\mathcal{R}$ by collecting elements of the trace $\pmb{1}$:
\bea
\mathcal{R}\equiv\pmb{1}\setminus\{r\}=\{1,2,\dots,r-1\}, \Label{Eq:RootSet}
\eea
where the last element $r\in \pmb{1}$ is always excluded\footnote{If a set $\pmb{B}$ is a subset of $\pmb{A}$, $\pmb{A}\setminus\pmb{B}$ is defined by moving all elements of $\pmb{B}$ from $\pmb{A}$.}.

{\noindent\bf{~Step-2}} Pick out the highest-weight element $\mathcal{H}_{\rho(l)}$  as well as other elements ${j_1},\dots,{j_u}$ (not necessary in the same relative order in $\mathsf{R}$) from the ordered set $\mathsf{R}$, then  construct a chain towards an element $w$ in the root set $\mathcal{R}$:
\bea
\mathcal{H}_{\rho(l)}\to j_u\to\dots\to j_1\to w. \Label{Eq:Chain1}
\eea
In the above chain, the $\mathcal{H}_{\rho(l)}$, ${j_1}$, ..., ${j_u}$ and $w$ are respectively mentioned as the \emph{starting element} (graviton or trace),  the \emph{internal elements} (gravitons and/or traces) and the \emph{ending element} (also mentioned as the root of the chain). The special case that a chain with no internal element $\mathcal{H}_{\rho(l)}\to w$ is allowed. The contribution of the chain (\ref{Eq:Chain1}) and the graphic expression are evaluated in the following way:
\begin{figure}
\centering
\includegraphics[width=0.65\textwidth]{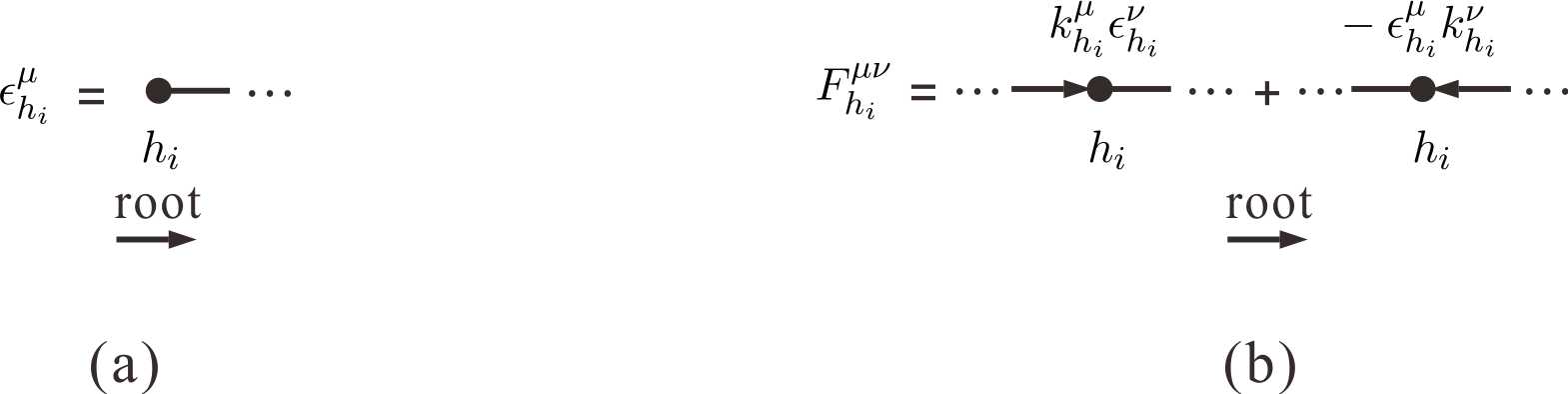}
\caption{The graphs for a starting graviton and an internal graviton are correspondingly given by (b).}\label{Fig:RefinedGraviton}
\end{figure}
\begin{figure}
\centering
\includegraphics[width=0.98\textwidth]{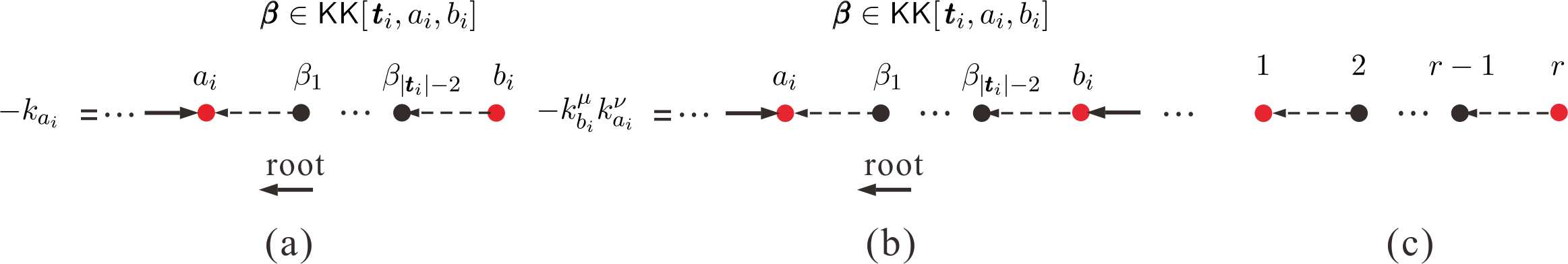}
\caption{The starting trace and an internal trace of a chain are respectively presented by the graphs (a) and (b). The trace $\pmb{1}\equiv\{1,2,\dots,r\}$ is expressed by the structure (c). In this paper, the two end nodes of a trace are always colored  red.}\label{Fig:RefinedTrace}
\end{figure}
 \begin{figure}
\centering
\includegraphics[width=0.55\textwidth]{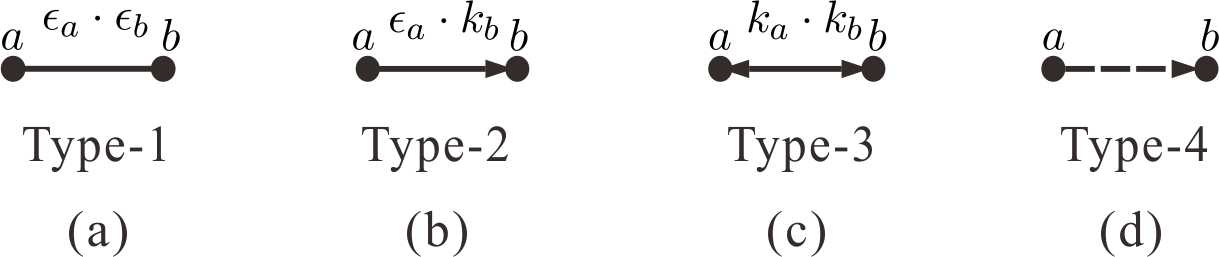}
\caption{Four types of lines in the refined graphic rule for multi-trace EYM amplitudes}\label{Fig:LineStyles}
\end{figure}
\begin{itemize}
\item {\bf Polarizations and momenta}~~Each half polarization $\epsilon^{\mu}$ of a graviton is expressed by a solid line connected to the graviton. The momentum $k^{\mu}$ of any node (graviton or gluon) is presented by a solid arrow line pointing to the node. Here, if an arrow points away from the direction of root, an extra minus should be dressed.

\item {\bf Starting and internal gravitons}~~If the starting element $\mathcal{H}_{\rho(l)}$ is a graviton $h_i$, it contributes a half polarization $\epsilon^{\mu}_{h_i}$ presented by the structure \figref{Fig:RefinedGraviton} (a). If an internal element is a graviton $h_i\in\{j_1,\dots,j_u\}$, it contributes a strength tensor $F^{\mu\,\nu}_{h_i}\equiv k^{\mu}_{h_i}\epsilon^{\nu}_{h_i}-k^{\nu}_{h_i}\epsilon^{\mu}_{h_i}$ and is expressed by the structure \figref{Fig:RefinedGraviton} (b).

\item {\bf Starting and internal traces}~~To express a gluon trace $\pmb{t}_i$, we select an ordered pair of gluons $\{a_i,b_i\}$ $(a_i,b_i\in\pmb{t}_i)$ where $a_i$ and $b_i$ play as the first and the last elements of the trace respectively.
    Then we arrange other gluons of $\pmb{t}_i$ in a relative order  $\pmb{\beta}\in \mathsf{KK}[\pmb{t}_i, a_i, b_i]$. Supposing that the trace can be written as $\pmb{t}_i=\{a_i,\pmb{X},b_i,\pmb{Y}\}$ where $\pmb{X}$ and $\pmb{Y}$ are the two ordered sets of gluons separated by $a_i$ and $b_i$, the $\mathsf{KK}[\pmb{t}_i, a_i, b_i]$ is defined by permutations $\pmb{X}\shuffle \pmb{Y}^{T}$ ($\pmb{Y}^{T}$ is the inverse order of $\pmb{Y}$ and $\pmb{A} \shuffle \pmb{B}$ for two ordered sets $\pmb{A}$, $\pmb{B}$ stands for the set of all permutations obtained by merging the two sets together with keeping the relative order of elements in each). Hence all gluons inside the trace are arranged in the relative order\footnote{Here, the $l$-th element $\beta(l)$ in the permutation $\pmb{\beta}$ is labeled by $\beta_l$ for short.}
    \bea
    a_i,\beta_1,\dots,\beta_{|\pmb{t}_i|-2},b_i, ~~(\{\beta_1,\dots,\beta_{|\pmb{t}_i|-2}\}\equiv \pmb{\beta}\in \mathsf{KK}[\pmb{t}_i, a_i, b_i]). \Label{Eq:PermutationTrace}
    \eea
     Supposing $a_i$ is nearer to root than $b_i$, we draw a dashed arrow line between any two adjacent gluons in the permutation (\ref{Eq:PermutationTrace}). Each arrow points towards the direction of the node $a_i$ as shown by \figref{Fig:RefinedTrace},  thus it also points towards the root.
    If the trace $\pmb{t}_i$ plays as the starting element $\mathcal{H}_{\rho(l)}$ of the chain (\ref{Eq:Chain1}), it is presented by the structure \figref{Fig:RefinedTrace} (a) and contributes a $-k_{a_i}^{\mu}$ to the coefficient. If $\pmb{t}_i\in\{j_1,\dots,j_u\}$ is an internal trace, it should be expressed by the structure \figref{Fig:RefinedTrace} (b) and contributes a $-k_{b_i}^{\mu}k_{a_i}^{\nu}$. As shown by \figref{Fig:RefinedTrace} (c), the special trace $\pmb{1}=\{1,2,\dots,r\}$ is presented via connecting adjacent gluons by dashed arrow lines whose arrows point towards the direction of the first gluon $1$.

\item {\bf Lorentz contractions and Line styles}~~Contracting the Lorentz indices accompanying with adjacent elements, we get the contribution of the  chain (\ref{Eq:Chain1})
       \bea
    \mathcal{E}_{\mathcal{H}_{\rho(l)}}\cdot \mathbb{F}_{j_u}\cdot...\cdot \mathbb{F}_{j_1}\cdot k_{w}, \Label{Eq:Chain2}
    \eea
in which
\bea
\mathcal{E}^{\mu}_{x}&=&\biggl\{
                \begin{array}{cc}
                  ~~\epsilon^{\mu}_{h_i}&\,(\text{if $x$ is a graviton $h_i$}) \\
                  ~~-k^{\mu}_{a_i} &\,\,\,\,\,\,\,(\text{if $x$ is a gluon trace $\pmb{t}_i$}) \\
                \end{array}\Label{Eq:EFY1}\\
\mathbb{F}^{\mu\,\nu}_{y}&=&\Biggl\{
                                                                                      \begin{array}{cc}
                                                                                        F^{\mu\,\nu}_{h_i}\equiv k^{\mu}_{h_i}\epsilon^{\nu}_{h_i}-k^{\nu}_{h_i}\epsilon^{\nu}_{h_i} & (\text{if $y$ is a graviton $h_i$}) \\
                                                                                        -k_{b_i}^{\mu}k_{a_i}^{\nu} &~\,\,\,\,(\text{if $y$ is a gluon trace $\pmb{t}_i$}) \\
                                                                                      \end{array}\Label{Eq:EFY2}.
                                                                                      \eea
 There are three types of lines \figref{Fig:LineStyles} (a), (b) and (c), which are resulted by the Lorentz contraction (\ref{Eq:Chain2}) and correspond to $\epsilon\cdot\epsilon$, $\epsilon\cdot k$ and $k\cdot k$. Recalling that dashed arrow lines \figref{Fig:LineStyles} (d) between gluons in a same trace have been introduced, we have four types of lines in all.
\end{itemize}
Redefine the ordered set $\mathsf{R}$ by removing the elements which have been used:
\bea
\mathsf{R}\to \mathsf{R}'=\mathsf{R}\setminus\{{j_1},\dots,{j_u},\mathcal{H}_{\rho(l)}\}\equiv\left\{\mathcal{H}_{\rho(1')},\mathcal{H}_{\rho(2')},\dots,\mathcal{H}_{\rho(l')}\right\}.\Label{Eq:RedefinedReferenceOrder}
\eea
and redefine the root set $\mathcal{R}$ by:
\bea
\mathcal{R}\to \mathcal{R}'=\mathcal{R}\cup {j_1}\cup\dots\cup {j_u}\cup\mathcal{H}_{\rho(l)}.\Label{Eq:RedefinedRootSet}
\eea
Here each element in $\mathcal{R}$ is either a graviton or a gluon (this  is different from the ordered set $\mathsf{R}$ where a trace is considered as a single element). If $j_i$  or $\mathcal{H}_{\rho(l)}$ in \eqref{Eq:RedefinedRootSet} is a graviton, it always stands for a single-element set $\{j_i\}$  or $\{\mathcal{H}_{\rho(l)}\}$.

{\noindent\bf{~Step-3}} Repeat the above step by using the new defined $\mathsf{R}$ and $\mathcal{R}$ until the ordered set $\mathsf{R}$ becomes empty. Then a fully connected tree graph $\mathcal{F}$ which is rooted at the gluon $1\in\pmb{1}$ is produced.


{\noindent\bf{~Step-4}} For a given graph $\mathcal{F}$, the coefficient $\mathcal{C}^{\mathcal{F}}$ in \eqref{Eq:PureYMExpansion} can be read off as the product of all factors corresponding to the type-1,-2 and -3 lines (see \figref{Fig:LineStyles}). The sign associated with such a graph gets two distinct contributions (i). $(-1)^{\mathcal{N}(\mathcal{F})}$, where $\mathcal{N}(\mathcal{F})$ is the number of arrows pointing away from the gluon  $1$; (ii). Each trace $\pmb{t}_i$ contributes a $(-1)^{|\pmb{t}_i,a_i,b_i|}$ for given $a_i$ and $b_i$, where $|\pmb{t}_i,a_i,b_i|$ is the number of elements in $\pmb{Y}_i$ if the trace $\pmb{t}_i$ can be written as $a_i, \pmb{X}_i, b_i, \pmb{Y}_i$.

{\noindent\bf{~Step-5}} Collect amplitudes $A(1,\pmb{\sigma}^{\mathcal{F}},r)$ for a given graph $\mathcal{F}$.~~In any graph $\mathcal{F}$,  the gluons $1$ (i.e. the root) and $r$ in the trace $\pmb{1}=\{1,2,\dots,r\}$ are always treated as the first and the last elements.  Permutations $\pmb{\sigma}^{\mathcal{F}}$ are determined as follows: {{(i)}.} Two adjacent nodes $x$ and $y$ which are connected by a line (of any style) must live on a path towards the root $1$. If $x$ is nearer to $1$ than $y$ on this path, we have $({\sigma}^{\mathcal{F}})^{-1}(x)<({\sigma}^{\mathcal{F}})^{-1}(y)$ where $({\sigma}^{\mathcal{F}})^{-1}(a)$ denotes the position\footnote{This is understood as follows: assuming that the position of $a$ in $\pmb{\sigma}^{\mathcal{F}}$ is $j$, we have $a={\sigma}^{\mathcal{F}}(j)$, hence it is reasonable to define $j=({\sigma}^{\mathcal{F}})^{-1}(a)$.} of $a$. {{(ii)}.} If there are several branches attached to a node, the relative order is defined by shuffling the branches together.

\begin{figure}
\centering
\includegraphics[width=0.96\textwidth]{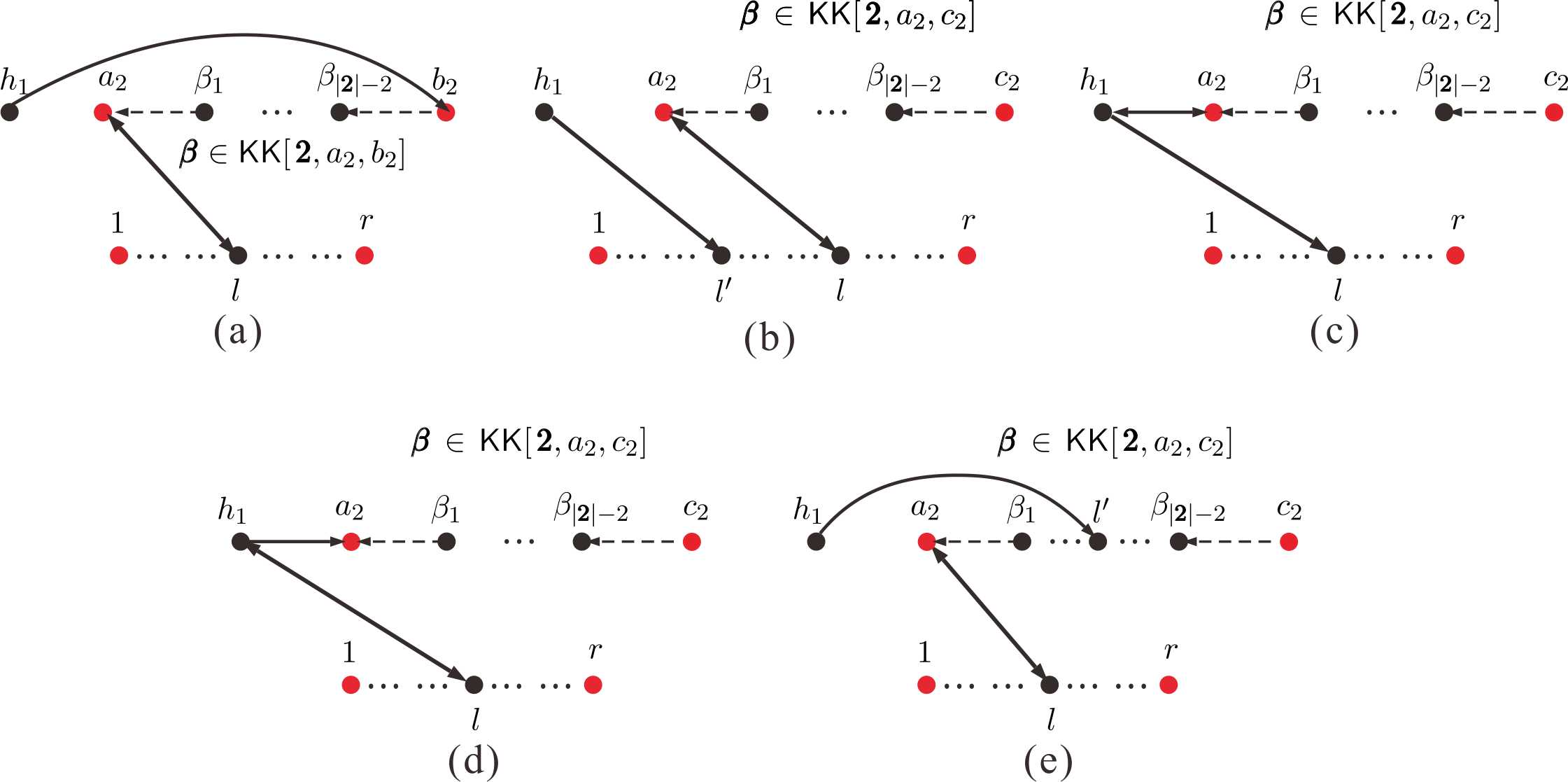}
\caption{Typical graphs for the pure YM expansion of the amplitude  $A(1,2\ldots r| \pmb{2}\Vert h_1)$. Graphs of structures (a), (b) and (c) contribute to the expansion with the reference order $\mathsf{R}=\left\{\pmb{2},h_1\right\}$. Graphs  (b), (c), (d) and (e) contribute to the expansion with the reference order $\mathsf{R}=\left\{h_1,\pmb{2}\right\}$. }\label{Fig:Example}
\end{figure}

When summing over all possible graphs constructed by the above steps, (i.e., {(i).} summing over all graphs with given $\{a_i,b_i\}$ pairs and given permutations $\pmb{\beta}\in \mathsf{KK}[\pmb{t}_i,a_i,b_i]$ for all traces, {(ii).} summing over all possible permutations $\pmb{\beta}\in \mathsf{KK}[\pmb{t}_i,a_i,b_i]$ for given $\{a_i,b_i\}$ pairs in all traces, {(iii).} summing over all possible choices of the $\{a_i,b_i\}$ pairs for internal traces and all possible choices of $a_i\in \pmb{t}_i, a_i\neq b_i$ for starting traces with fixed $b_i$'s\footnote{Notice that the refined graphic rule is given by applying the recursive expansion (\ref{Eq:MultiTrace}) iteratively. The $b_i$ of the starting trace in each step of recursive expansion can be chosen freely (see \secref{sec:Review}). This implies that the $b_i$'s for the same trace, which plays as the starting element of different chains, can be chosen differently. In this paper, we fix the $b_i\in\pmb{t}_i$ as the same one in all the graphs where the trace $\pmb{t}_i$ plays as a starting element.
}), we finally  arrive the expansion  (\ref{Eq:PureYMExpansion}). In the coming subsection, we show a concrete example to explain this rule.

\subsection{Examples for the refined graphic rule}
Now we take the double-trace amplitude  $A(1,2\ldots r| \pmb{2}\Vert h_1)$  with one graviton as an example. For this amplitude, the reference order can be chosen as either $\mathsf{R}=\left\{\pmb{2},h_1\right\}$ or $\mathsf{R}=\left\{h_1,\pmb{2}\right\}$ in which the highest-weight element is respectively the graviton $h_1$ or the gluon trace $\pmb{2}$. We study these two cases separately.

If $\mathsf{R}=\left\{\pmb{2},h_1\right\}$, the typical graphs are shown by \figref{Fig:Example} (a), (b) and (c). Correspondingly, these graphs contribute
    \bea
   \mathcal{C}^{\text{(a)}}&=&-(\epsilon_{h_1}\cdot k_{b_2})(k_{a_2}\cdot k_l),\nn
   \pmb{\sigma}^{\text{(a)}}&\in&\{2,\dots,l,\{l+1,\dots,r-1\}\shuffle\{a_2,\beta_1,\dots,\beta_{|\pmb{2}|-2},b_2,h_1\}\},\Label{Eq:EGa}\\
   \mathcal{C}^{\text{(b)}}&=&-(\epsilon_{h_1}\cdot k_{l'})(k_{a_2}\cdot k_{l}),\nn \pmb{\sigma}^{\text{(b)}}&\in&\bigl\{2,\dots,l',\{h_1\}\shuffle\bigl\{l'+1,\dots,l,\{l+1,\dots,r-1\}\shuffle\{a_2,\beta_1,\dots,\beta_{|\pmb{2}|-2},c_2\}\bigr\}\bigr\},\Label{Eq:EGb}\\
 \mathcal{C}^{\text{(c)}}&=&-(\epsilon_{h_1}\cdot k_{l})(k_{a_2}\cdot k_{h_1}),\nn \pmb{\sigma}^{\text{(c)}}&\in&\{2,\dots,l,\{l+1,\dots,r-1\}\shuffle\{h_1,a_2,\beta_1,\dots,\beta_{|\pmb{2}|-2},c_2\}\},\Label{Eq:EGc}
    \eea
    where the permutations $\pmb{\beta}=\{\beta_1,\beta_2,\dots,\beta_{|\pmb{2}|-2}\}$ in $\pmb{\sigma}^{\text{(a)}}$ and $\pmb{\sigma}^{\text{(b)},\text{(c)}}$ satisfy $\pmb{\beta}\in \mathsf{KK}[\pmb{2},a_2,b_2]$ and $\pmb{\beta}\in \mathsf{KK}[\pmb{2},a_2,c_2]$ respectively. The signs caused by arrows pointing away from the root are absorbed into the coefficients $\mathcal{C}$ in eqs. (\ref{Eq:EGa}), (\ref{Eq:EGb}) and (\ref{Eq:EGc}), but the signs  $(-1)^{|\pmb{t}_2,a_2,b_2|}$, $(-1)^{|\pmb{t}_2,a_2,c_2|}$ induced by the trace are not. Then the sum over all graphs in \eqref{Eq:PureYMExpansion} for the amplitude $A(1,2\ldots r| \pmb{2}\Vert h_1)$  is given by
    \\
    \\
     \bea
     &&\Sl_{l\in\{1,\dots,r-1\}}\,\Biggl[\,\widetilde{\Sl_{\{a_2,b_2\}\subset\pmb{2}}}\,\Sl_{\pmb{\sigma}^{\text{(a)}}}\, \mathcal{C}^{\text{(a)}}A(1,\pmb{\sigma}^{\text{(a)}},r)\Label{Eq:RefinedEG1}\\
     &&~~~~~~~~~~~~~~~~~~~~~~~~~~~~~+\widetilde{\Sl_{\scriptsize{\substack{a_2\in\pmb{2}\\a_2\neq c_2(c_2\in \pmb{2})}}}}\biggl(\,\Sl_{l'\in\{1,\dots,r-1\}}\Sl_{\pmb{\sigma}^{\text{(b)}}} \mathcal{C}^{\text{(b)}}A(1,\pmb{\sigma}^{\text{(b)}},r)+ \Sl_{\pmb{\sigma}^{\text{(c)}}}\,\mathcal{C}^{\text{(c)}}\,A(1,\pmb{\sigma}^{\text{(c)}},r)\biggr)\Biggr].\nonumber
     \eea
     where the summation over $\pmb{\beta}\in \mathsf{KK}[\pmb{2},a_2,b_2]$ ($\pmb{\beta}\in \mathsf{KK}[\pmb{2},a_2,c_2]$) and the factor $(-1)^{|\pmb{t}_2,a_2,b_2|}$ ($(-1)^{|\pmb{t}_2,a_2,c_2|}$) corresponding to the trace $\pmb{2}$ have been absorbed into the notation $\widetilde{\sum}_{\{a_2,b_2\}\subset\pmb{2}}$ ($\widetilde{\sum}_{\scriptsize{\substack{a_2\in\pmb{2}\\a_2\neq b_2(b_2\in \pmb{2})}}}$), i.e.,
     \bea
     \widetilde{\Sl_{\{a_2,b_2\}\subset\pmb{2}}}&\equiv&\Sl_{\{a_2,b_2\}\subset\pmb{2}}(-1)^{|\pmb{t}_2,a_2,b_2|}\Sl_{\pmb{\beta}\in \mathsf{KK}[\pmb{2},a_2,b_2]},\nn
     ~~\widetilde{\Sl_{\scriptsize{\substack{a_2\in\pmb{2}\\a_2\neq c_2(c_2\in \pmb{2})}}}}&\equiv&{\Sl_{\scriptsize{\substack{a_2\in\pmb{2}\\a_2\neq c_2(c_2\in \pmb{2})}}}}(-1)^{|\pmb{t}_2,a_2,c_2|}\Sl_{\pmb{\beta}\in \mathsf{KK}[\pmb{2},a_2,c_2]}.\Label{Eq:TildeSummation}
     \eea
      The $c_2\in\pmb{2}$ in the second term of \eqref{Eq:RefinedEG1} is arbitrarily fixed because the trace $\pmb{2}$ plays as the starting trace in \figref{Fig:Example} (b) and (c).

\begin{figure}
\centering
\includegraphics[width=0.96\textwidth]{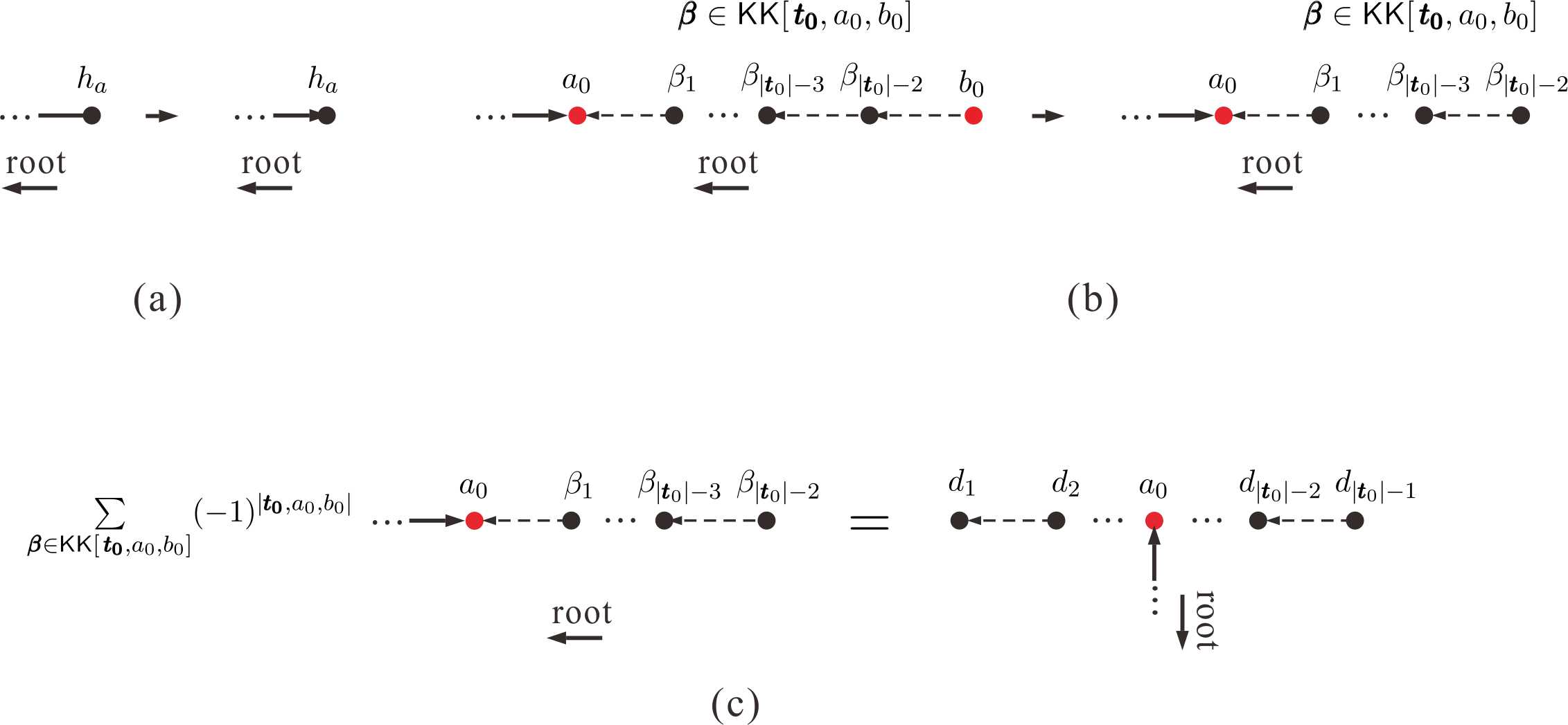}
\caption{The induced identities (\ref{Eq:InducedID1}) and (\ref{Eq:InducedID2}) are obtained respectively by  imposing the replacements (a) and (b) on the graphs in \eqref{Eq:PureYMExpansion} where the corresponding highest-weight elements are the graviton $h_a$ and the trace $\pmb{t}_0$.
The sum over all $\pmb{\beta}\in \mathsf{KK}[\pmb{t}_0,a_0,b_0]$ (with the sign $(-1)^{|\pmb{t}_0,a_0,b_0|}$) on the RHS of the replacement (b) can be further expressed by (c), where the full trace $\pmb{t}_0$ is supposed to be  $d_1,d_2,\dots d_{j-1},a_0=d_j,d_{j+1},\dots,d_{|\pmb{t}_0|-1},b_0=d_{|\pmb{t}_0|}$. }\label{Fig:Replacement}
\end{figure}

 If $\mathsf{R}=\left\{h_1,\pmb{2}\right\}$, the typical graphs are given by \figref{Fig:Example} (b), (c), (d) and (e). The expression of \figref{Fig:Example} (b) and (c) are already shown by \eqref{Eq:EGb} and \eqref{Eq:EGc}. The contributions from \figref{Fig:Example} (d) and (e) read
    \bea
   \mathcal{C}^{\text{(d)}}&=&(k_{h_1}\cdot k_{l})(\epsilon_{h_1}\cdot k_{a_2}), \nn \pmb{\sigma}^{\text{(d)}}&\in&\{2,\dots,l,\{l+1,\dots,r-1\}\shuffle\{h_1,a_2,\beta_1,\dots,\beta_{|\pmb{2}|-2},c_2\}\}\Label{Eq:EGd}\\
   \mathcal{C}^{\text{(e)}}&=&-(\epsilon_{h_1}\cdot k_{l'})(k_{a_2}\cdot k_{l}),\nn \pmb{\sigma}^{\text{(e)}}&\in&\bigl\{2,\dots,l,\{l+1,\dots,r-1\}\shuffle\{a_2,\beta_1,\dots,l'=\beta_j,\{\beta_{j+1},\dots,\beta_{|\pmb{2}|-2},c_2\}\shuffle\{h_1\}\}\bigr\},\Label{Eq:EGe}
    \eea
where $l'$ can be any element in the trace $\pmb{2}$. The sum over all graphs in \eqref{Eq:PureYMExpansion} for the amplitude $A(1,2\ldots r| \pmb{2}\Vert h_1)$  is then written as
\bea
&&\Sl_{l\in\{1,\dots,r-1\}}\widetilde{\Sl_{\scriptsize{\substack{a_2\in\pmb{2}\\a_2\neq c_2(c_2\in \pmb{2})}}}}\Biggl[\,\Sl_{l'\in\{1,\dots,r-1\}} \Sl_{\pmb{\sigma}^{\text{(b)}}}\mathcal{C}^{\text{(b)}}A(1,\pmb{\sigma}^{\text{(b)}},r)+\,\Sl_{\pmb{\sigma}^{\text{(c)}}}\mathcal{C}^{\text{(c)}}A(1,\pmb{\sigma}^{\text{(c)}},r)\Label{Eq:RefinedEG2}\\
&&~~~~~~~~~~~~~~~~~~~~~~~~~~~~~~~~~~~~~~~~~~~~~~~~~~~~~~~~~~~~~+\Sl_{\pmb{\sigma}^{\text{(d)}}}\mathcal{C}^{\text{(d)}}A(1,\pmb{\sigma}^{\text{(d)}},r)+\Sl_{l'\in\pmb{2}} \Sl_{\pmb{\sigma}^{\text{(e)}}}\mathcal{C}^{\text{(e)}}A(1,\pmb{\sigma}^{\text{(e)}},r)\Biggr].\nonumber
\eea

\subsection{Induced identities by refined graphs}\label{sec:InducedIDBYRefinedGF}
As pointed in \cite{Du:2017gnh}, one can induce a nontrivial identity (see \eqref{Eq:Type-IID} or \eqref{Eq:Type-IIID}) of EYM amplitudes by imposing each of the following conditions on the recursive expansion (\ref{Eq:MultiTrace}): (i). the gauge invariance condition of a graviton, (ii). the cyclic symmetry of a gluon trace. When the EYM amplitudes on the LHS of \eqref{Eq:Type-IID} and \eqref{Eq:Type-IIID} are further expanded repeatedly according to \eqref{Eq:MultiTrace}, we finally arrive identities for pure YM amplitudes. Such an identity can also be obtained by imposing the condition (i) or (ii) on \eqref{Eq:PureYMExpansion} straightforwardly. As shown in the appendix, we only need to study identities which are induced by conditions of the highest-weight element. Those identities induced by conditions of elements other than the highest-weight one are essentially treated as cases with a smaller $\pmb{\mathcal{H}}$ set.

If the highest-weight element $\mathcal{H}_{\rho(l)}$ in the reference order (\ref{Eq:ReferenceOrder}) is a graviton $h_a$, the gauge invariance condition of $h_a$ states that the expansion (\ref{Eq:PureYMExpansion}) vanishes when  $\epsilon^{\mu}_{h_a}$ is replaced by $k^{\mu}_{h_a}$:
    \bea
\boxed{0=\Sl_{\mathcal{F}} \mathcal{C}^{\mathcal{F}}\big |_{\epsilon_{h_a}\to k_{h_a}}\biggl[\,\Sl_{\pmb{\sigma}^{\mathcal{F}}}A(1,\pmb{\sigma}^{\mathcal{F}},r)\,\biggr]}.\Label{Eq:InducedID1}
    \eea
    The corresponding graphs can be obtained by the replacement \figref{Fig:Replacement} (a) \footnote{Although, the arrow in \figref{Fig:Replacement} (a) points away from the root, we do not associate a minus sign with this graph, for an overall sign in the identity \eqref{Eq:InducedID1} can always be neglected.}.

 If the highest-weight element $\mathcal{H}_{\rho(l)}$ in the reference order (\ref{Eq:ReferenceOrder}) is a gluon trace $\pmb{t}_0$, the following identity is induced by the replacement
 \bea
 \{a_0,\mathsf{KK}[\pmb{t}_0,a_0,b_0],b_0\}\to\{a_0,\mathsf{KK}[\pmb{t}_0,a_0,b_0]\},\Label{Eq:Repacement}
 \eea
  for any $a_0\in\pmb{t_0}~(a_0\neq b_0)$:
    \bea
    \boxed{0=\Sl_{\mathcal{F}} \mathcal{C}^{\mathcal{F}}\biggl[\,\Sl_{\pmb{\sigma}^{\mathcal{F}}}A(1,\pmb{\sigma}^{\mathcal{F}},r)\,\biggr]\bigg |_{\substack{\{a_0,\mathsf{KK}[\pmb{t}_0,a_0,b_0],b_0\}\\\to \{a_0,\mathsf{KK}[\pmb{t}_0,a_0,b_0]\}}}}.\Label{Eq:InducedID2}
    \eea
    This identity is understood as follows:  when we consider $\{a_0,\mathsf{KK}[\pmb{t}_0,a_0,b_0]\}$ (for $a_0\in\pmb{t}_0$) (which does not contain the fixed gluon $b_0\in\pmb{t}_0$) as the highest-weight element $\mathcal{H}_{\rho(l)}$ (see \figref{Fig:Replacement} (b)) and then apply the refined graphic rule, the total contribution of the RHS of \eqref{Eq:PureYMExpansion} must vanish. As stated in \cite{Du:2017gnh}, this identity is essentially a result of the cyclic symmetry of the trace $\pmb{t}_0$. It is worth pointing out an interesting property \cite{Du:2017gnh}: when we sum over $\pmb{\beta}\in \mathsf{KK}[\pmb{t}_0,a_0,b_0]$ with the sign $(-1)^{|\pmb{t}_0,a_0,b_0|}$, the trace after the replacement \figref{Fig:Replacement} (b) becomes the RHS of \figref{Fig:Replacement} (c) (where  $\pmb{t}_0$ is supposed to be $d_1,d_2,\dots d_{j-1},a_0=d_j,d_{j+1},\dots,d_{|\pmb{t}_0|-1},b_0=d_{|\pmb{t}|_0}$). In the coming sections, we always use the RHS of \figref{Fig:Replacement} (c) to stand for the highest-weight element in the identity (\ref{Eq:InducedID2}).

In the next section, we show that the identities (\ref{Eq:InducedID1}) and (\ref{Eq:InducedID2}) induced from the double-trace EYM amplitude $A(1,\cdots,r|\pmb{2}\Vert h_1)$  can be expanded in terms of graph-based BCJ relations.

\section{Identities induced from $A(1,\cdots,r|\pmb{2}\Vert h_1)$ as combinations of BCJ relations}\label{Sec:Examples}
We now take the identities induced from the amplitude $A(1,\cdots,r|\pmb{2}\Vert h_1)$  as explicit examples and show that   \eqref{Eq:InducedID1} and \eqref{Eq:InducedID2}  can be written as combinations of graph-based BCJ relations (\ref{Eq:Graph-based-BCJ}) (thus traditional BCJ relations (\ref{Eq:BCJRelation})). Critical features of these examples are further summarized.
\begin{figure}
	\centering
	\includegraphics[width=0.97\textwidth]{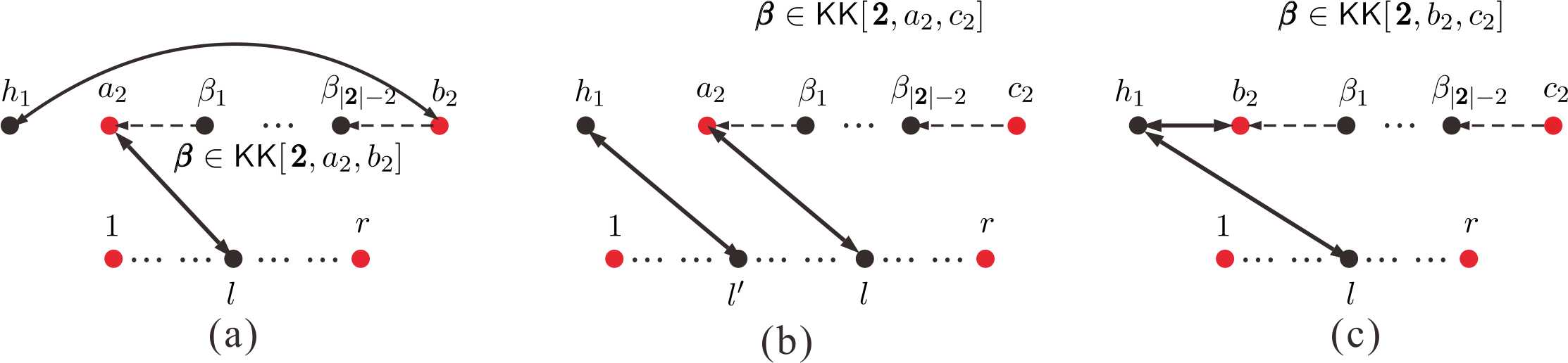}
	\caption{Typical graphs for the identity induced from the double-trace amplitude $A(1,2,\cdots,r|\pmb{2}\Vert h_1)$ with the reference order $\mathsf{R}=\{\pmb{2},h_1\}$. The trace $\pmb{2}$ plays as an internal trace in (a) and a starting trace in (b), (c).}\label{Fig:Example1}
\end{figure}
\subsection{Example-1: the identity (\ref{Eq:InducedID1}) induced from $A(1,2,\cdots,r|\pmb{2}\Vert h_1)$}\label{sec:Example-1}
When the reference order is chosen as $\mathsf{R}=\{\pmb{2},h_1\}$, the highest-weight element is the graviton $h_1$. Thus the expansion (\ref{Eq:PureYMExpansion}) for the double-trace amplitude $A(1,2,\cdots,r|\pmb{2}\Vert h_1)$ induces an identity (\ref{Eq:InducedID1}) under the replacement $\epsilon^{\mu}_{h_1}\to k^{\mu}_{h_1}$. Typical graphs for this identity are given by \figref{Fig:Example1} (a), (b) and (c) which are correspondingly  obtained from \figref{Fig:Example} (a), (b) and (c) via the replacement \figref{Fig:Replacement} (a). The total contributions of these graphs are then written as
\bea
T^{\text{(a)}}&=&\Sl_{\pmb{\sigma}^{\text{(a)}}}(k_{h_1}\cdot k_{b_2})(-k_{a_2}\cdot k_l)A\big(1,\,\pmb{\sigma}^{\text{(a)}},\,r\big),\Label{Eq:EG1a}\\
T^{\text{(b)}}&=&\Sl_{\pmb{\sigma}^{\text{(b)}}}(k_{h_1}\cdot k_{l'})(-k_{a_2}\cdot k_{l})\,A\big(1,\,\pmb{\sigma}^{\text{(b)}},\,r\big),\Label{Eq:EG1b}\\
T^{\text{(c)}}&=&\Sl_{\pmb{\sigma}^{\text{(c)}}}(k_{h_1}\cdot k_{l})(-k_{b_2}\cdot k_{h_1})A\big(1,\,\pmb{\sigma}^{\text{(c)}},\,r\big),\Label{Eq:EG1c}
\eea
where $\pmb{\sigma}^{\text{(a)}}$, $\pmb{\sigma}^{\text{(b)}}$ and $\pmb{\sigma}^{\text{(c)}}$ are already presented  in  \eqref{Eq:EGa}, \eqref{Eq:EGb} and \eqref{Eq:EGc} respectively. Then the full expression of the RHS of the identity (\ref{Eq:InducedID1}) induced from $A(1,\cdots,r|\pmb{2}\Vert h_1)$ is given by
\bea
\Biggl[\,\Sl_{l\in\{1,\dots,r-1\}}\widetilde{\Sl_{{\scriptsize\{a_2,b_2\}\subset\pmb{2}}}}T^{\text{(a)}}\Biggr]+\Biggl[\,\Sl_{l,l'\in\{1,\dots,r-1\}}\widetilde{\Sl_{\scriptsize\substack{a_2\in\pmb{2}\\a_2\neq c_2\,(c_2\in\pmb{2})}}}T^{\text{(b)}}\Biggr]+\Biggl[\,\Sl_{l\in\{1,\dots,r-1\}}\widetilde{\Sl_{\scriptsize\substack{b_2\in\pmb{2}\\b_2\neq c_2\,(c_2\in\pmb{2})}}}T^{\text{(c)}}\Biggr],\Label{Eg:SumOfT}
\eea
where a summation notation with a tile is already defined by  \eqref{Eq:TildeSummation}.
%

To investigate the relationship between the induced identity and BCJ relations, we define the {\bf\emph{standard basis set}} for a gluon trace $\pmb{i}$ by the set of permutations $\bigl\{\{b_i,\pmb{\beta}\in\mathsf{KK}[\,\pmb{i},b_i,c_i],c_i\}\big|b_i\neq c_i, b_i\in\pmb{i}\bigr\}$ for an arbitrarily fixed end node $c_i\in\pmb{i}$ (i.e. we do not sum over the end node $c_i$). Graphically, any two adjacent nodes for a given permutation in the standard basis set $\{b_i,\pmb{\beta}\in\mathsf{KK}[\,\pmb{i},b_i,c_i],c_i\}$ are connected by a dashed arrow line pointing towards the node $b_i$. A starting trace defined by the refined graphic rule is already expressed by the standard basis because an end node of this trace is already fixed. Moreover, all internal traces can be expanded by the standard basis, according to the following nontrivial property:
\bea
	\boxed{\centering
	\includegraphics[width=0.91\textwidth]{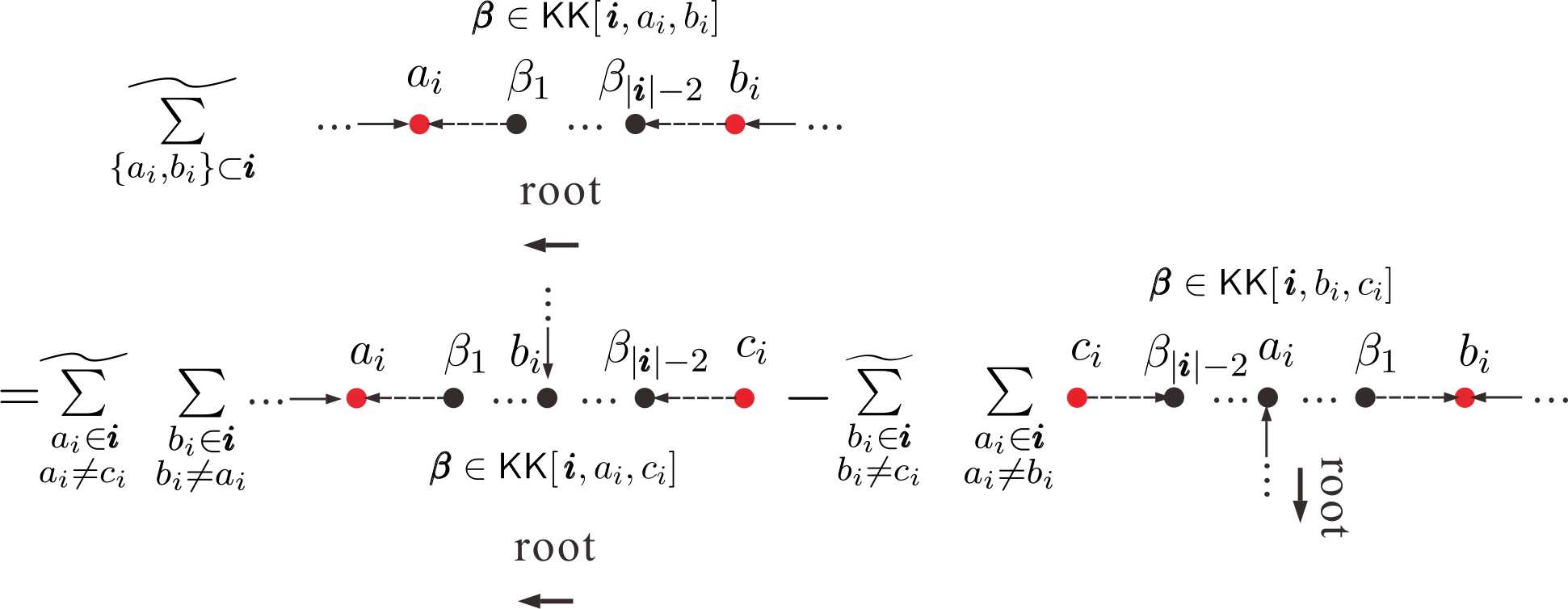}},\nn\Label{Eq:SplittingTraces}
\eea
where each graph stands for its full contribution (including coefficients and amplitudes).  The LHS of the above equation is an internal trace structure defined by the refined graphic rule.  On the RHS, the trace $\pmb{i}$ is expressed by the standard basis with one end $c_i\in\pmb{i}$ fixed. The node $a_i$ ($b_i$) on both sides of \eqref{Eq:SplittingTraces} must be connected to a same node outside the trace $\pmb{i}$ via the same type of line (i.e.,  type-2 or type-3 line). The summation notations with a tilde in \eqref{Eq:SplittingTraces} was already defined by \eqref{Eq:TildeSummation}.  If a dashed arrow points away from the root, an extra minus should be dressed. We leave the proof of \eqref{Eq:SplittingTraces} in \appref{sec:Appendix}.

\begin{figure}
	\centering
	\includegraphics[width=1\textwidth]{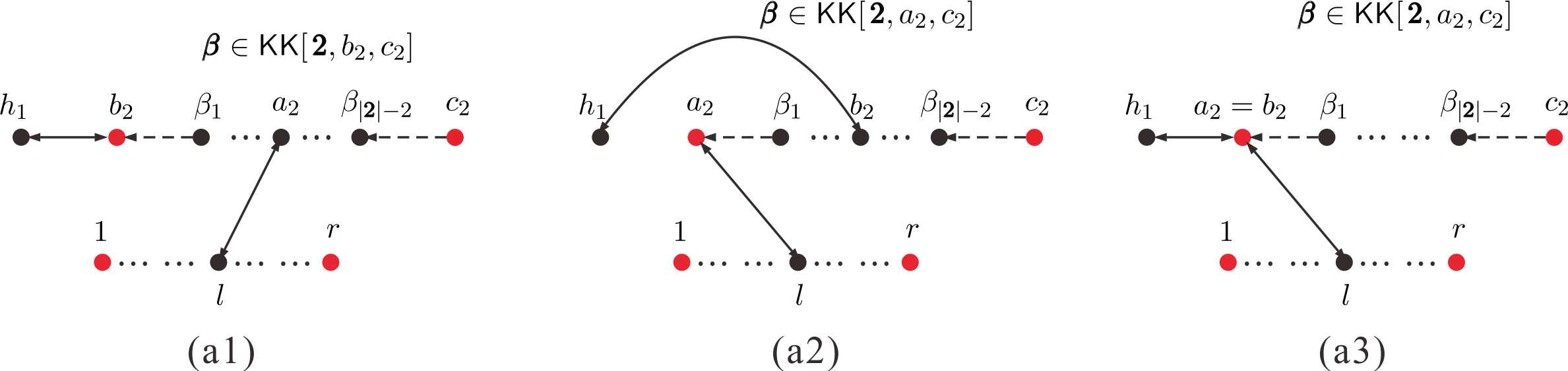}
	\caption{When the identity (\ref{Eq:SplittingTraces}) is applied, the contribution of \figref{Fig:Example1} (a) splits into (a1) and (a2) which are expressed by standard basis. The graph (a1) comes from the second term of \eqref{Eq:SplittingTraces}, thus it must be associated with an extra minus. The graph (a3) is a spurious graph which cannot be directly obtained by the refined graphic rule and the identity (\ref{Eq:SplittingTraces}). It can be considered as the $a_i=b_i$ supplement to both terms on the RHS of \eqref{Eq:SplittingTraces}. }\label{Fig:Splitting2}
\end{figure}

Now we apply the relation (\ref{Eq:SplittingTraces}) to the first term of \eqref{Eg:SumOfT} for a given $l$. Since the $c_i$ in \eqref{Eq:SplittingTraces} can be chosen arbitrarily, we just choose the $c_i\in \pmb{i}$ (in this example $\pmb{i}=\pmb{2}$)  as  the fixed gluon $c_2$ in the trace $\pmb{2}$ of \figref{Fig:Example1} (b) and (c) for convenience. Then the summation $\widetilde{\sum}_{{\scriptsize\{a_2,b_2\}\subset\pmb{2}}}T^{\text{(a)}}$ in the first term of \eqref{Eg:SumOfT} splits into
\bea
\widetilde{\Sl_{{\scriptsize\{a_2,b_2\}\subset\pmb{2}}}}T^{\text{(a)}}=-\widetilde{\Sl_{{\scriptsize \substack{b_2\in\pmb{2}\\b_2\neq c_2}}}}\Sl_{\scriptsize\substack{a_2\in\pmb{2}\\a_2\neq b_2}}T^{\text{(a1)}}+\widetilde{\Sl_{{\scriptsize\substack{a_2\in\pmb{2}\\a_2\neq c_2}}}}\Sl_{\scriptsize\substack{b_2\in\pmb{2}\\b_2\neq a_2}}T^{\text{(a2)}},\Label{Eg:SumOfT1}
\eea
where $T^{\text{(a1)}}$ and $T^{\text{(a2)}}$ are corresponding to \figref{Fig:Splitting2} (a1) and (a2).
Substituting \eqref{Eg:SumOfT1} into \eqref{Eg:SumOfT} and introducing the contribution of \emph{spurious graph} \figref{Fig:Splitting2} (a3) by $0=T^{\text{(a3)}}-T^{\text{(a3)}}$,
we rewrite \eqref{Eg:SumOfT} as the sum of $I_1$ and $I_2$ which are respectively defined by
\bea
I_1\equiv\widetilde{\Sl_{{\scriptsize \substack{b_2\in\pmb{2}\\b_2\neq c_2}}}}\biggl[\,\Sl_{l\in\{1,\dots,r-1\}}\biggl(\,-\Sl_{\scriptsize\substack{a_2\in\pmb{2}\\a_2\neq b_2}}T^{\text{(a1)}}-T^{\text{(a3)}}|_{a_2=b_2}+T^{\text{(c)}}\biggr)\biggr]\Label{Eq:I1}
\eea
and
\bea
I_2\equiv\widetilde{\Sl_{\scriptsize\substack{a_2\in\pmb{2}\\a_2\neq c_2}}}\biggl[\,\Sl_{l\in\{1,\dots,r-1\}}\biggl(\Sl_{\scriptsize\substack{b_2\in\pmb{2}\\b_2\neq a_2}}T^{\text{(a2)}}+T^{\text{(a3)}}+\Sl_{l'\in\{1,\dots,r-1\}}T^{\text{(b)}}\biggr)\biggr].\Label{Eq:I2}
\eea
In the following, we analyze $I_1$ and $I_2$ in turn and prove that both of them can be expanded in terms of graph-based BCJ relations.

\begin{figure}
	\centering
	\includegraphics[width=0.4\textwidth]{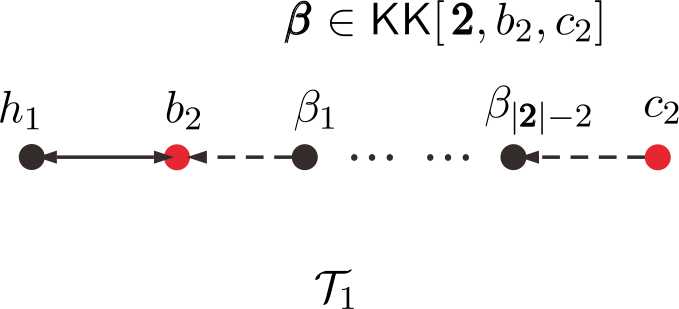}
	\caption{The graphs \figref{Fig:Splitting2} (a1), (a3) and \figref{Fig:Example1} (c) can be reproduced by connecting the structure $\mathcal{T}_1$ with a gluon $l\in\{1,2,\dots,r-1\}=\pmb{1}\setminus \{r\}$ via a type-3 line. This further implies that the $I_1$ in \eqref{Eq:I1} can be expanded by graph-based BCJ relations (\ref{Eq:Graph-based-BCJ}) with $\mathcal{T}=\mathcal{T}_1$. }\label{Fig:T1}
\end{figure}

{\bf (i).} It is easy to see the graphs \figref{Fig:Splitting2} (a1), (a3) and \figref{Fig:Example1} (c) corresponding to the three terms of \eqref{Eq:I1} can be reproduced by drawing a type-3 line between a gluon $l\in\{1,2,\dots,r-1\}=\pmb{1}\setminus \{r\}$ and a node $a\in \mathcal{T}_1$ where $\mathcal{T}_1$ is the tree structure \figref{Fig:T1}. Particularly, $a$ is given by $a_2\in\pmb{2}~(a_2\neq b_2)$ for  \figref{Fig:Splitting2} (a1),  $b_2$ for \figref{Fig:Splitting2} (a3) and  $h_1$ for \figref{Fig:Example1} (c). The kinematic factors of \figref{Fig:Splitting2} (a1), (a3) and \figref{Fig:Example1} (c) can then be uniformly given by $(k_{h_1}\cdot k_{b_2})(k_{a}\cdot k_l)$. Let us count the sign: Given $b_2\in \pmb{2}\,(b_2\neq c_2)$ in \eqref{Eq:I1}, there is an overall sign $(-1)^{|\pmb{2},b_2,c_2|}$ which has already been absorbed into the summation notation with a tilde $\widetilde{\sum}_{{\scriptsize \substack{b_2\in\pmb{2}\\b_2\neq c_2}}}$. The sign for each term inside the square brackets is collected as follows: (1). According to the refined graphic rule, any graph $\mathcal{F}$ with $\mathcal{N}(\mathcal{F})$ arrows (for both solid and dashed arrow lines) pointing away from the root $1$ is associated with a sign $(-1)^{\mathcal{N}(\mathcal{F})}$. (2). Each of (a1) and (a3) has an extra sign $(-1)$. The above observations further lead to the following pattern:
    \begin{itemize}
    \item  Once $a\in \mathcal{T}_1$ has been chosen, the sign is independent of the choice of $l\in\{1,2,\dots,r-1\}=\pmb{1}\setminus \{r\}$ because neither  $\mathcal{N}(\mathcal{F})$ nor the extra sign in \eqref{Eq:I1} relies on $l$. On another hand, two graphs with adjacent $a\in \mathcal{T}_1$ are associated with opposite signs.

    \item All permutations established by the graph with any given $a\in \mathcal{T}_1$ have the form
    \bea
    \{1,\pmb{\sigma}\in\{2,\dots,r-1\}\shuffle\mathcal{T}_1|_{a},r\}.\Label{Eq:PermutationsEG1}
     \eea
     Here $\mathcal{T}_1|_{a}$ is introduced as the relative orders between nodes of the tree $\mathcal{T}_1$ when the node $a$ is considered as the leftmost one. According to the refined graphic rule, when we connect a type-3 line between $a\in \mathcal{T}_1$ and $l\in \pmb{1}\setminus\{r\}$, $l$ must be nearer to the root $1$ than $a$. In other words, for a given  permutation $\pmb{\sigma}$ in \eqref{Eq:PermutationsEG1}, $l\in\{1,\dots,r-1\}$ can be any node satisfying $\sigma^{-1}(l)<\sigma^{-1}(a)$. Then the total coefficient (which comes from the type-3 line between $a$ and $l$) for the permutation $\pmb{\sigma}$ is collected as
    $-k_a\cdot Y_a(\pmb{\sigma})$ where $Y^{\mu}_a(\pmb{\sigma})\equiv \sum_{\scriptsize{\sigma^{-1}(l)<\sigma^{-1}(a)}} k^{\mu}_l$ (the momentum of the root $1$ is always included in this summation).

    \end{itemize}
Altogether,  $I_1$ in \eqref{Eq:I1} can be reexpressed by
    \bea
    I_1=\widetilde{\Sl_{{\scriptsize \substack{b_2\in\pmb{2}\\b_2\neq c_2}}}}(k_{h_1}\cdot k_{b_2})(-)^{\mathcal{F}({x_0})}\biggl[\,\Sl_{a\in \mathcal{T}_1}f^a\Sl_{\zeta\in\mathcal{T}_1|_{a}}\Sl_{\pmb{\sigma}\in\zeta\shuffle\{2,\dots,r-1\}}(k_a\cdot Y_a(\pmb{\sigma}))A(1,\pmb{\sigma},r)\biggr],
    \eea
where $(-)^{\mathcal{F}({x_0})}$ denotes the sign for the graph $\mathcal{F}$ with $a=x_0$ ($x_0\in\mathcal{T}_1$). The $f^{a}$ for any $a\in\mathcal{T}_1$ is fixed as (i). $f^{x_0}=1$, (ii). $f^{x_1}=-f^{x_2}$ if $x_1$ and $x_2$ are two adjacent nodes in $\mathcal{T}_1$. Therefore, the expression in the square brackets is just the LHS of the graph-based BCJ relation (\ref{Eq:Graph-based-BCJ}) which has been proven to be a combination of traditional BCJ relations (\ref{Eq:BCJRelation}) (see \cite{Hou:2018bwm}).

{\bf (ii).} For $I_2$, all the permutations established by the graphs \figref{Fig:Example1} (b), \figref{Fig:Splitting2} (a2) and (a3) have the form
$\{1, \pmb{\sigma}\in\{h_1\}\shuffle\,\pmb{\gamma}, r\}$, where
\bea
\pmb{\gamma}\in\{2,\dots,l,\{l+1,\dots,r-1\}\shuffle\{a_2,\mathsf{KK}[\,\pmb{2},a_2,c_2],c_2\}\}.\Label{Eq:gamma}
\eea
Coefficient for each permutation $\pmb{\sigma}\in\{h_1\}\shuffle\,\pmb{\gamma}$ is collected as $(-k_{a_2}\cdot k_l)(k_{h_1}\cdot Y_{h_1}(\pmb{\sigma}))$. Hence $I_2$ turns to
\bea
I_2=\widetilde{\Sl_{\scriptsize\substack{a_2\in\pmb{2}\\a_2\neq c_2}}}\,\Sl_{l\in\{1,\dots,r-1\}}(-k_{a_2}\cdot k_l)\Sl_{\pmb{\gamma}}\biggl[\,\Sl_{\pmb{\sigma}\in\{h_1\}\shuffle\,\pmb{\gamma}}(k_{h_1}\cdot Y_{h_1}(\pmb{\sigma}))A(1,\pmb{\sigma},r)\biggr],
\eea
in which $\pmb{\gamma}$ satisfy \eqref{Eq:gamma}. Apparently, the expression in the square brackets is just the LHS of a special case of  BCJ relation (\ref{Eq:BCJRelation}), which can also be understood as the graph-based BCJ relation (\ref{Eq:Graph-based-BCJ}) when the tree graph $\mathcal{T}$ is the single node $h_1$.

\begin{figure}
	\centering
	\includegraphics[width=1\textwidth]{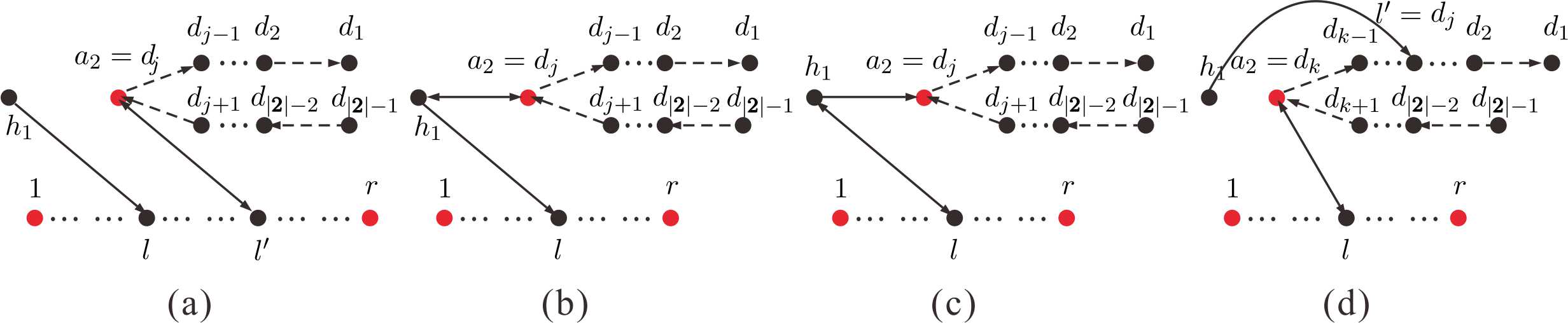}
	\caption{Typical graphs for the identity (\ref{Eq:InducedID2}) induced from the double-trace amplitude $A(1,2,\cdots,r|\pmb{2}\Vert h_1)$ with the reference order $\mathsf{R}=\{h_1,\pmb{2}\}$. Here, gluons of the trace $\pmb{2}$ are supposed to be in the cyclic order $d_1,d_2,\dots,d_{|\pmb{2}|}$ and the gluon $d_{|\pmb{2}|}$ is the removed gluon $b_0$ in the induced identity (\ref{Eq:InducedID2}). }\label{Fig:Example2}
\end{figure}
%

\subsection{Example-2: the  identity (\ref{Eq:InducedID2}) induced from $A(1,2,\cdots,r|\pmb{2}\Vert h_1)$}\label{sec:Example-2}

When the reference order for the expansion (\ref{Eq:PureYMExpansion}) of the double-trace EYM amplitude $A(1,2,\cdots,r|\pmb{2}\Vert h_1)$ is chosen as $\mathsf{R}=\{h_1,\pmb{2}\}$ (i,e., the trace $\pmb{2}$ is the highest-weight element), the cyclic symmetry of the trace $\pmb{2}$ induces the identity (\ref{Eq:InducedID2}) where typical graphs are presented as \figref{Fig:Example} (b)-(e). According to the discussions in \secref{sec:InducedIDBYRefinedGF}, we do the replacement \figref{Fig:Replacement} (b) and then \figref{Fig:Replacement} (c) on the graphs \figref{Fig:Example} (b)-(e). Thus graphs \figref{Fig:Example2} (a)-(d) with the following contributions are obtained correspondingly:
\bea
T^{\text{(a)}}&=&\Sl_{\pmb{\sigma}^{\text{(a)}}}(\epsilon_{h_1}\cdot k_{l})(-k_{a_2}\cdot k_{l'})\,A(1,\pmb{\sigma}^{\text{(a)}},r),~~~~\,~~~~T^{\text{(b)}}=\Sl_{\pmb{\sigma}^{\text{(b)}}}(\epsilon_{h_1}\cdot k_l)(-k_{a_2}\cdot k_{h_1})A(1,\pmb{\sigma}^{\text{(b)}},r),\nn
T^{\text{(c)}}&=&\Sl_{\pmb{\sigma}^{\text{(c)}}}(-\epsilon_{h_1}\cdot k_{a_2})(-k_{h_1}\cdot k_{l})A(1,\pmb{\sigma}^{\text{(c)}},r),~~~~~\,T^{\text{(d)}}=\Sl_{\pmb{\sigma}^{\text{(d)}}}(\epsilon_{h_1}\cdot k_{l'})(-k_{a_2}\cdot k_{l})\,A(1,\pmb{\sigma}^{\text{(d)}},r),
\eea
in which,
\bea
 \pmb{\sigma}^{\text{(a)}}&\in&\bigl\{2,\dots,l,\{h_1\}\nn
&&~~~~~~~~~~~~~\shuffle\bigl\{l+1,\dots,l',\{l'+1,\dots,r-1\}\shuffle \{a_2=d_j,\{d_{j+1},\dots,d_{|\pmb{2}|-1}\}\shuffle\{d_{j-1},\dots,d_1\}\}\bigr\}\bigr\},\nn
 \pmb{\sigma}^{\text{(b),(c)}}&\in&\{2,\dots,l,\{l+1,\dots,r-1\}\shuffle\{h_1,a_2=d_j,\{d_{j+1},\dots,d_{|\pmb{2}|-1}\}\shuffle\{d_{j-1},\dots,d_1\}\},\nn
\pmb{\sigma}^{\text{(d)}}&\in&\bigl\{2,\dots,l,\{l+1,\dots,r-1\}\nn
&&~~~~~~~~~~~~~\shuffle\{a_2=d_k,\{d_{k+1},\dots,d_{|\pmb{2}|-1}\}\shuffle\{d_{k-1},\dots,l'=d_{j},\{h_1\}\shuffle\{d_{j-1},\dots,d_1\}\}\}\bigr\}.
\eea
Once all graphs are summed over, we arrive the RHS of the induced identity \eqref{Eq:InducedID2} for the amplitude $A(1,2,\cdots,r|\pmb{2}\Vert h_1)$:
\bea
\underbrace{\Sl_{l,l'\in\{1,\dots,r-1\}}\Sl_{j=1}^{|\pmb{2}|-1}T^{\text{(a)}}+\Sl_{l\in\{1,\dots,r-1\}}\Sl_{j=1}^{|\pmb{2}|-1}\Bigl(T^{\text{(b)}}}_{I_1}+\underbrace{T^{\text{(c)}}\Bigr)+\Sl_{l\in\{1,\dots,r-1\}}\Sl_{j,k=1}^{|\pmb{2}|-1}T^{\text{(d)}}}_{I_2},\Label{Eq:EgII}
\eea
where contributions of all graphs of the form \figref{Fig:Example2} (a), (b) and (c), (d) were collected as $I_1$ and $I_2$ respectively.
Now we prove that both $I_1$ and $I_2$ in \eqref{Eq:EgII} can be expanded in terms of BCJ relations.
\begin{figure}
	\centering
	\includegraphics[width=0.65\textwidth]{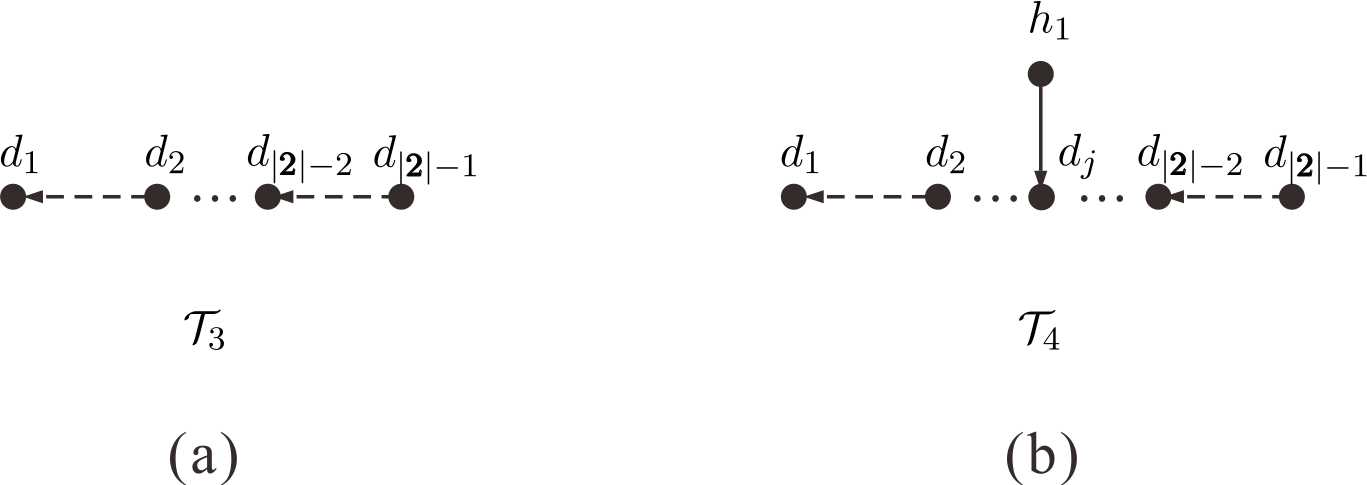}
	\caption{The terms $I_1$ and $I_2$ in \eqref{Eq:EgII} can be expanded in terms of graph-based BCJ relations (\ref{Eq:Graph-based-BCJ}), where the corresponding tree graphs $\mathcal{T}$ are chosen as the graph (a) and the graph (b).}\label{Fig:T3T4}
\end{figure}

For the $I_1$ part in \eqref{Eq:EgII}, the summation over $j=1,\dots,|\pmb{2}|-1$ is nothing but just the summation over all nodes $a_2\in\mathcal{T}_3$ where $\mathcal{T}_3$ is the tree graph \figref{Fig:T3T4} (a). All permutations established by  the graphs \figref{Fig:Example2} (a), (b) with a given $a_2\in \mathcal{T}_3$ and a given $l\in\{1,\dots,r-1\}$ have the form
\bea
\big\{1,\pmb{\sigma}\in (\mathcal{T}_3|_{a_2})\shuffle \pmb{\gamma},r\big\}.
 \eea
Here, $\pmb{\gamma}\in\{2,\dots,l,\{h_1\}\shuffle\{l+1,\dots,r-1\}\}$ and $\mathcal{T}_3|_{a_2}$ denotes the relative orders of nodes in $\mathcal{T}_3$  when $a_2$ is considered as the leftmost one. Based on a similar discussion with the example-1 in \secref{sec:Example-1}, we find the following patterns: (1). The coefficients $(k_{a_2}\cdot k_{l'})$ ($l'\in\{1,\dots,r-1\}$ for \figref{Fig:Example2} (a) and $l'=h_1$ for \figref{Fig:Example2} (b)) corresponding to a same $\pmb{\sigma}\in (\mathcal{T}_3|_{a_2})\shuffle \pmb{\gamma}$ with different choices of $l'$ are collected as $(-k_{a_2}\cdot Y_{a_2}(\pmb{\sigma}))$; (2). Any two graphs, where $a_2\in \mathcal{T}_3$ are chosen as adjacent nodes, have opposite signs. Then the $I_1$ part in \eqref{Eq:EgII} is expressed by
\bea
I_1&=&\Sl_{l\in\{1,\dots,r-1\}}\Sl_{\small\substack{\pmb{\gamma}\in\{2,\dots,l,\{h_1\}\\\shuffle\{l+1,\dots,r-1\}\}}}(\epsilon_{h_1}\cdot k_l)(-)^{\mathcal{F}({x_0})}\nn
&&~~~~~~~~~~~~~~~~~~~~~~~~~~~~~~~~~~~~~~~\times\biggl[\,\Sl_{a_2\in \mathcal{T}_3}f^{a_2}\Sl_{\pmb{\sigma}\in\{\mathcal{T}_3|_{a_2}\shuffle\,\pmb{\gamma}\}}(-k_{a_2}\cdot Y_{a_2}(\pmb{\sigma}))A(1,\pmb{\sigma},r)\biggr].
\eea
Here $(-)^{\mathcal{F}(x_0)}$ is the sign for a graph with $a_2=x_0\in \mathcal{T}_3$. Since $(-)^{\mathcal{F}({x_0})}$  has been extracted as an overall sign, we have $f^{x_0}=1$.  If $x_1, x_2\in \mathcal{T}_3$ are two adjacent nodes, we have $f^{x_1}=-f^{x_2}$. Obviously, the expression in the square brackets is nothing but (up to a total minus) the LHS of a graph-based BCJ relation (\ref{Eq:Graph-based-BCJ}). As a result, $I_1$ is a combination of traditional BCJ relations.

The $I_2$ part in \eqref{Eq:EgII} can be analyzed following a parallel discussion with $I_1$ but replacing the tree graph $\mathcal{T}_3$ by $\mathcal{T}_4$ (see \figref{Fig:T3T4} (b)) for $d_j\in \{d_1,\dots,d_{|\pmb{2}|-1}\}$ and replacing $\pmb{\gamma}$ by $\{2,\dots,r-1\}$:
    \bea
I_2=\Sl_{j=1}^{|\pmb{2}|-1}(\epsilon_{h_1}\cdot k_{d_j}) (-)^{\mathcal{F}({x_0})}\biggl[\,\Sl_{a_2\in \mathcal{T}_4}f^{a_2}\Sl_{\pmb{\sigma}\in\{\mathcal{T}_4|_{a_2}\shuffle\,\{2,\dots,r-1\}\}}(-k_{a_2}\cdot Y_{a_2}(\pmb{\sigma}))A(1,\pmb{\sigma},r)\biggr],
\eea
where $x_0\in \mathcal{T}_4$ and $f^{x_0}=1$. Again, the $f^{a_2}$'s for adjacent choices of $a_2$ have the opposite signs.  Up to a total sign, the expression in the square brackets is just the LHS of graph-based BCJ relation (\ref{Eq:Graph-based-BCJ}), where the tree graph $\mathcal{T}_4$  for a given $j$ is \figref{Fig:T3T4} (b). Thus, we conclude that $I_2$ is a combination of BCJ relations.

\subsection{Common features of the examples}\label{sec:CommentsExamples}
Now let us extract some common features from the examples, which will be extended to general cases in the next section.

\begin{figure}
	\centering
	\includegraphics[width=1\textwidth]{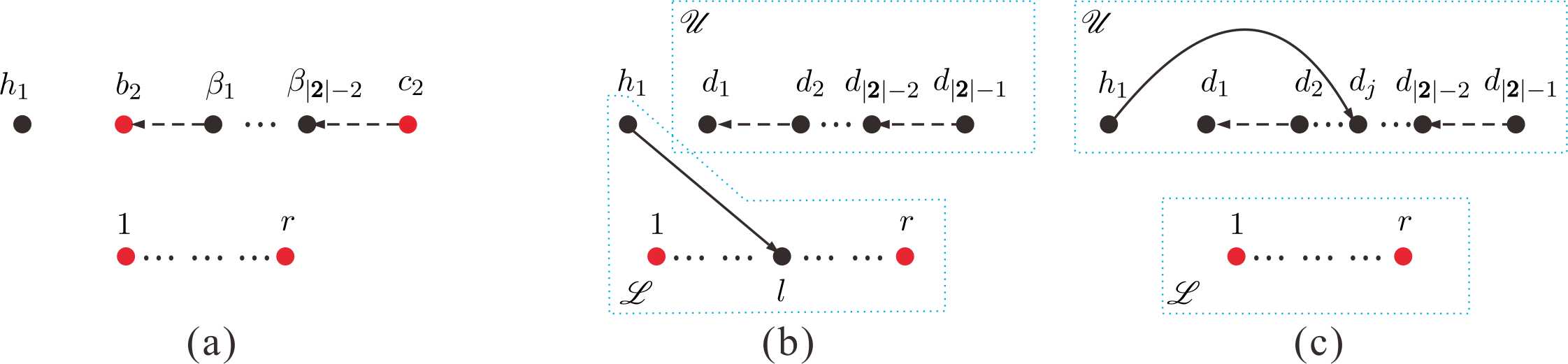}
	\caption{A typical skeleton in example-1 is presented as the graph (a) which consists of three components. Graphs (b) and (c) are possible structures of skeletons in example-2. The subgraphs $\mathscr{U}$ and $\mathscr{L}$ in each of (b) and (c) are correspondingly the final upper and lower blocks.}\label{Fig:SkeletonEG}
\end{figure}

\noindent{~\bf(i). Expressing traces by standard basis}~~
In example-1, the trace $\pmb{2}$ played as an internal trace in \figref{Fig:Example1} (a) and a starting trace in either \figref{Fig:Example1} (b) or (c). In the latter cases, the trace $\pmb{2}$ was already expressed by the standard basis, i.e., one end of the trace, the gluon $c_2$, was fixed. In the former case, both ends of the trace $\pmb{2}$ were not fixed (in other words both are summed over). In order to expand the trace $\pmb{2}$ in \figref{Fig:Example1} (a) by the standard basis, we have made used of the splitting trace relation (\ref{Eq:SplittingTraces}), in which the fixed node was conveniently chosen as the same element (i.e. $c_2$) with that in \figref{Fig:Example1} (b) and (c).

\noindent{~\bf(ii). Skeletons and components}~~We define {\emph{skeletons}} by removing all type-3 lines from the graphs where all traces, except the highest-weight element in $\mathsf{R}$ (if it is a trace), are already expressed by the standard basis. Since each graph defined by the refined graphic rule is a connected tree graph, its skeleton must be a disconnected graph. Each maximally connected subgraph of a skeleton is called a {\emph{component}}. A typical skeleton in example-1 is given by \figref{Fig:SkeletonEG} (a) (for $b_2\in \pmb{2}$, $b_2\neq c_2$) which consists of three components. Skeletons in example-2 have two distinct structures, \figref{Fig:SkeletonEG} (b) and (c), each of which has two components. From the examples, we can see \emph{any skeleton must  have at least two components that involve the highest-weight element (graviton or trace) and the trace $\pmb{1}$ respectively}.

\begin{figure}
	\centering
	\includegraphics[width=0.67\textwidth]{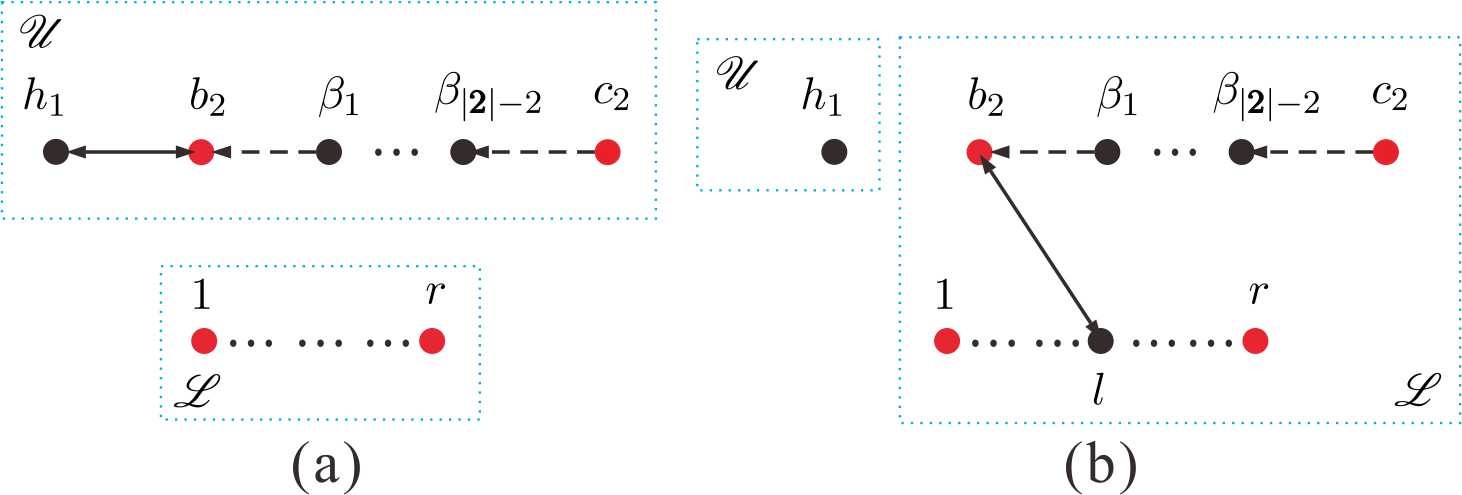}
	\caption{Typical configurations of the final upper and lower blocks $\mathscr{U}$, $\mathscr{L}$ for the skeleton \figref{Fig:SkeletonEG} (a).}\label{Fig:FinalUL1}
\end{figure}

\noindent{~\bf(iii). The final upper and lower blocks}~~Any graph in the examples can be reproduced by connecting a type-3 line between \emph{the final upper and lower blocks} which are two mutually disjoint connected subgraphs and respectively contain the highest-weight element and the trace $\pmb{1}$. In example-2, each of the skeletons \figref{Fig:SkeletonEG} (b) and (c) already consists of only two disjoint connected subgraphs $\mathscr{U}$ and $\mathscr{L}$ which serve as the final upper and lower blocks.  In example-1, there are three components in the skeleton \figref{Fig:SkeletonEG} (a). A typical configuration of the final upper and lower blocks is constructed when we connect  $b_2$  in \figref{Fig:SkeletonEG} (a) to either (i). $h_1$ (see \figref{Fig:FinalUL1} (a)) or (ii). an element in $\{1,\dots,r-1\}$ via a type-3 line (see \figref{Fig:FinalUL1} (b)).

\noindent{~\bf(iv). Physical and spurious graphs}~~For a given configuration of the final upper and lower blocks $\mathscr{U}$ and $\mathscr{L}$, we can connect two nodes $x\in\mathscr{U}$ and $y\in\mathscr{L}\setminus\{r\}$ (recalling that the gluon $r$ is always excluded) via a type-3 line.  Then a fully connected graph is constructed. In example-2, the graphs \figref{Fig:Example2} (a), (b) (and \figref{Fig:Example2} (c), (d)) are reproduced by connecting the final upper and lower blocks in  \figref{Fig:SkeletonEG} (b) (and \figref{Fig:SkeletonEG} (c)) via a type-3 line. Similarly, the graphs \figref{Fig:Splitting2} (a1), \figref{Fig:Example1} (c) (and \figref{Fig:Splitting2} (a2), \figref{Fig:Example1} (b)) in example-1 are constructed from \figref{Fig:FinalUL1} (a) (and (b)). All the graphs \figref{Fig:Example1} (b), (c), \figref{Fig:Splitting2} (a1), (a2) and \figref{Fig:Example2} (a)-(d) are graphs in standard basis which are directly defined by the refined graphic rule. These graphs are called \emph{physical graphs}. The \emph{spurious graph} \figref{Fig:Splitting2} (a3), which is not defined by the refined graphic rule, can be reproduced from either \figref{Fig:FinalUL1} (a) (connecting $b_2\in \mathscr{U}$ with $l\in \mathscr{L}\setminus\{r\}$) or \figref{Fig:FinalUL1} (b) (connecting $h_1\in \mathscr{U}$ with $b_2\in \mathscr{L}\setminus\{r\}$). In the former case,  a minus sign is introduced so that the spurious graph constructed by distinct ways cancel with one another. Therefore, for any skeleton, the sum of all physical graphs  can be given by {\bf(1).} summing over all possible configurations of the final upper and lower blocks $\mathscr{U}$, $\mathscr{L}$, {\bf (2).} for a given $\mathscr{U}$ and $\mathscr{L}$, connecting two nodes $x\in\mathscr{U}$ and $y\in\mathscr{L}\setminus \{r\}$ via a type-3 line and summing over all possible choices of $x$, $y$ (in other words summing over all possible physical and spurious graphs corresponding to $\mathscr{U}$ and $\mathscr{L}$).

\noindent{~\bf(v). Induced identities as combinations of graph-based BCJ relations}~~A crucial observation is that the sum over all the graphs corresponding to a given configuration of the final upper and lower blocks $\mathscr{U}$ and $\mathscr{L}$ (i.e. either $I_1$ or $I_2$ in each example) is a combination of graph-based BCJ relations (\ref{Eq:Graph-based-BCJ}).

In the next section, we extend these observations to general cases and show that both identities (\ref{Eq:InducedID1}) and (\ref{Eq:InducedID2}) can be expanded in terms of BCJ relations.

\section{General study}\label{sec:GeneralStudy}
To investigate the general induced identities (\ref{Eq:InducedID1}) and (\ref{Eq:InducedID2}) in a unified way, we write them as
\bea
0=\Sl_{\mathcal{G}}\mathcal{C}^{\mathcal{G}}\,\Sl_{\pmb{\sigma}^{\mathcal{G}}}\,A(1,\pmb{\sigma}^{\mathcal{G}},r),~\Label{Eq:InducedIDUnified}
\eea
where the graphs $\mathcal{G}$ are obtained by imposing the replacement \figref{Fig:Replacement} (a) or (b), which corresponds to \eqref{Eq:InducedID1} or \eqref{Eq:InducedID2}, on the graphs $\mathcal{F}$ in \eqref{Eq:PureYMExpansion}.

When we introduce skeletons $\mathcal{G}'$ by deleting all type-3 lines from the graphs $\mathcal{G}$ and expressing all traces by standard basis according to \eqref{Eq:SplittingTraces}, \eqref{Eq:InducedIDUnified} is rearranged as
\bea
0=\widetilde{\Sl_{\mathcal{G}'}}\mathcal{P}^{[\mathcal{G}']}\left[\,\Sl_{\mathcal{G}\supset \mathcal{G}' }\,\Sl_{\pmb{\sigma}^{\mathcal{G}}}(-)^{\mathcal{G}}\mathcal{K}^{[\mathcal{G}\setminus\mathcal{G}']}A(1,\pmb{\sigma}^{\mathcal{G}},r)\right].~\Label{Eq:InducedIDUnified2}
\eea
In the above equation, the summation notation $\widetilde{\sum}_{\mathcal{G}'}$ means that all possible skeletons $\mathcal{G}'$  are summed over and the signs $(-1)^{|\pmb{t}_i,a_i,c_i|}$ and/or $(-1)^{|\pmb{t}_i,b_i,c_i|}$ accompanying to the traces in each $\mathcal{G}'$  are absorbed. The factor $\mathcal{P}^{[\mathcal{G}']}$ denotes the kinematic factor corresponding to the skeleton $\mathcal{G}'$. Since a skeleton does not involve any type-3 line, $\mathcal{P}^{[\mathcal{G}']}$ only consists of $\epsilon\cdot\epsilon$ and $\epsilon\cdot k$ factors. In the expression inside the square brackets, all possible physical graphs $\mathcal{G}$ (i.e. graphs generated by refined graphic rule with all internal traces expressed by the standard basis) containing the skeleton $\mathcal{G}'$ and all permutations $\pmb{\sigma}^{\mathcal{G}}$ for each graph $\mathcal{G}$ are summed over. The factor $\mathcal{K}^{[\mathcal{G}\setminus\mathcal{G}']}$ in \eqref{Eq:InducedIDUnified2} stands for the product of all $k\cdot k$ factors that are presented by the type-3 lines in $\mathcal{G}$.  Those signs caused by arrows pointing away from the root and the extra signs caused by the second term of \eqref{Eq:SplittingTraces} are all collected as $(-)^{\mathcal{G}}$.
\begin{figure}
	\centering
	\includegraphics[width=0.87\textwidth]{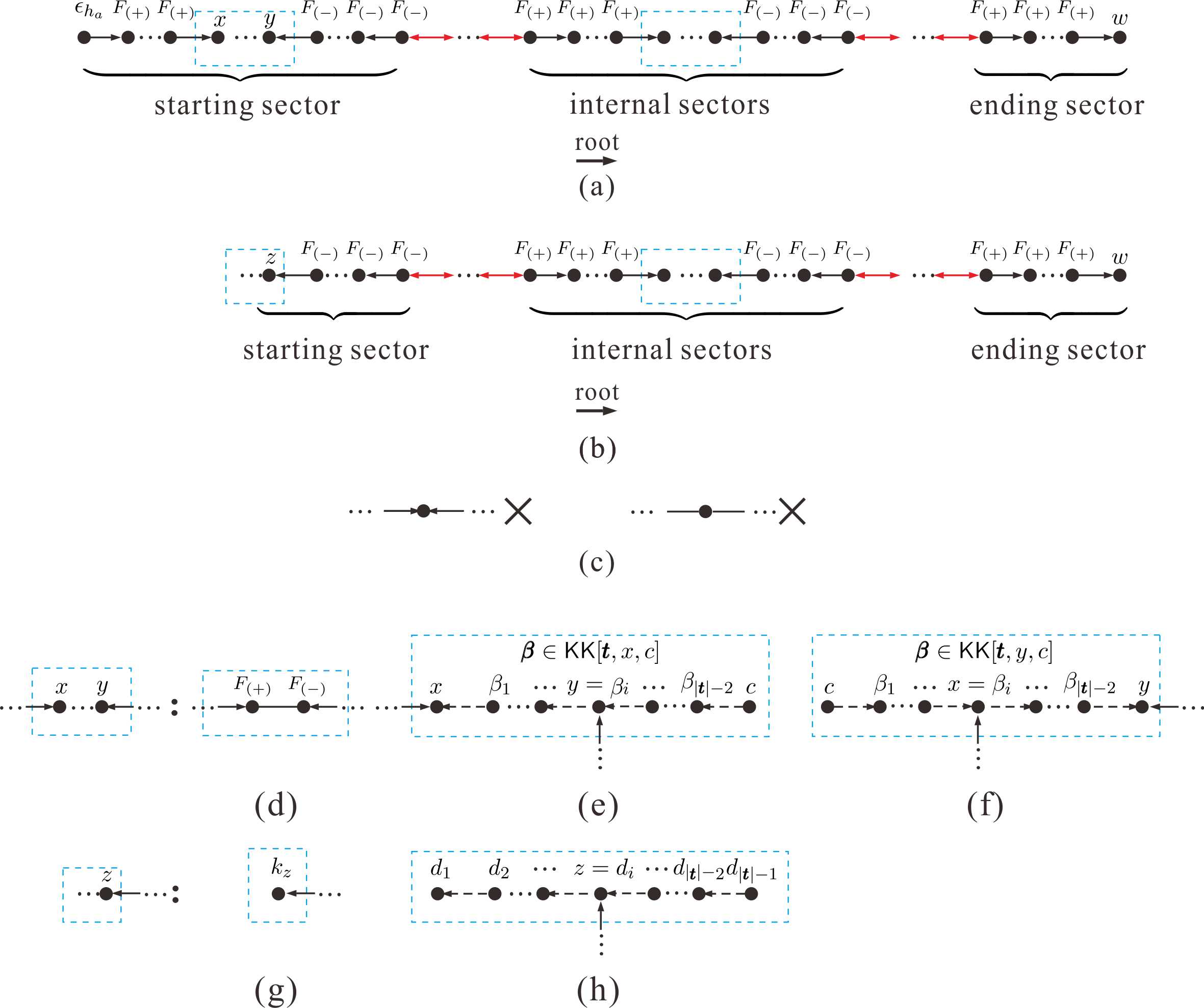}
	\caption{The graph (a) is a chain that does not involve the highest-weight element $\mathcal{H}_{\rho(l)}$ in the reference order $\mathsf{R}$. The graph (b) is a chain involving the highest-weight element. Each node (except for the ending node $w$) outside the boxed structures in the chains (a) and (b) is a graviton and we define $F^{\mu\nu}_{(+)}\equiv k^{\mu}\epsilon^{\nu}$, $F^{\mu\nu}_{(-)}\equiv -\epsilon^{\nu}k^{\mu}$. The structures shown by the graph (c) are not allowed. Each boxed structure in the chain (a) and in an internal sector of the chain (b) can be (1). two gravitons connected together by a type-1 line (as shown by the graph (d)) or (2). a gluon trace in standard basis (as shown by (e) and (f) which correspond to the two terms in \eqref{Eq:SplittingTraces}). The node $c$ in (e), (f) denotes the fixed node of a trace $\pmb{t}$.  The boxed structure in the starting sector of (b) can be either a graviton (as shown by (g)) or  a gluon trace  $\pmb{t}\to\{d_1,d_2,\dots,d_{|\pmb{t}|-1}\}$ (as shown by (h)).  }\label{Fig:ChainStructure1}
\end{figure}

As observed in \secref{sec:CommentsExamples},  all physical graphs $\mathcal{G}$ involving a given skeleton $\mathcal{G}'$  can be generated by connecting the components via type-3 lines in a proper way: (i). first generate all possible configurations of the final upper and lower blocks $\mathscr{U}\oplus\mathscr{L}$; (ii). then connect a type-3 line between two nodes $x\in\mathscr{U}$ and $y\in \mathscr{L}\setminus\{r\}$ appropriately.
One should take care of the step (ii) because spurious graphs may also be produced. Nevertheless, the spurious graphs in fact all cancel out in the examples. Hence we suppose that the expression in the square brackets in \eqref{Eq:InducedIDUnified2}  can be generally written as
\bea
I[\mathcal{G}']\equiv \Sl_{\mathscr{U}\oplus\mathscr{L}}\mathcal{K}^{[\mathscr{U}\oplus\mathscr{L}\setminus\mathcal{G}']}\Biggl[\,
\Sl_{\substack{x\in\mathscr{U}\\y\in\mathscr{L}\setminus\{r\}}}\,\Sl_{\pmb{\sigma}^{\mathcal{G}}}(-)^{\mathcal{G}}\,(k_x\cdot k_y)\,A(1,\pmb{\sigma}^{\mathcal{G}},r)\Biggr].~\Label{Eq:InducedIDUnified3}
\eea
Here, $\mathcal{K}^{[\mathscr{U}\oplus\mathscr{L}\setminus\mathcal{G}']}$ is the product of all $k\cdot k$ factors corresponding to the given configuration of the final upper and lower blocks $\mathscr{U}$, $\mathscr{L}$, while $k_x\cdot k_y$ is the factor corresponding to the type-3 line between $\mathscr{U}$ and $\mathscr{L}$. The first summation in \eqref{Eq:InducedIDUnified3} is taken over all possible configurations of the final upper and lower blocks $\mathscr{U}$, $\mathscr{L}$ for the skeleton $\mathcal{G}'$.  In the square brackets, all choices of nodes $x\in \mathscr{U}$ and $y\in\mathscr{L}\setminus\{r\}$ as well as the permutations $\pmb{\sigma}^{\mathcal{G}}$  defined by the (physical or spurious) graph $\mathcal{G}$ (determined by $\mathscr{U}$, $\mathscr{L}$, $x$ and $y$) are summed over. The sign for the (physical or spurious) graph $\mathcal{G}$ is denoted by $(-)^{\mathcal{G}}$.


In this section, we study \eqref{Eq:InducedIDUnified3} schematically. We first classify components of skeletons, then show how to construct the final upper and lower blocks from a given skeleton $\mathcal{G}'$. After that, we show all spurious graphs cancel out. Thus the summation over all possible choices of  $x\in\mathscr{U}$ and $ y\in\mathscr{L}\setminus\{r\}$  is equivalent to summing over all possible physical graphs for  given $\mathscr{U}$ and $\mathscr{L}$. At last, we demonstrate that the expression inside the square brackets in \eqref{Eq:InducedIDUnified3} is  a combination of BCJ relations.

%
%
%
%
\begin{figure}
	\centering
	\includegraphics[width=0.8\textwidth]{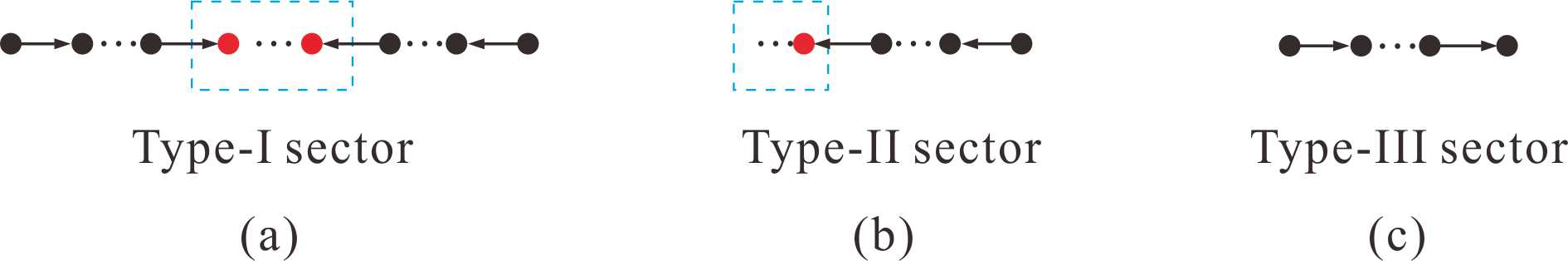}
	\caption{There are three types of sectors (a), (b) and (c). The starting and internal sectors of the chain \figref{Fig:ChainStructure1} (a) and the internal sectors of the chain \figref{Fig:ChainStructure1} (b) have the same general structure (a). Such sectors are called type-I sectors. The starting sector of  the chain \figref{Fig:ChainStructure1} (b) is called the type-II sector and has the pattern (b). The ending sector of any chain has the structure (c) and is called a type-III sector. }\label{Fig:ChainStructure3}
\end{figure}
\begin{figure}
	\centering
	\includegraphics[width=0.7\textwidth]{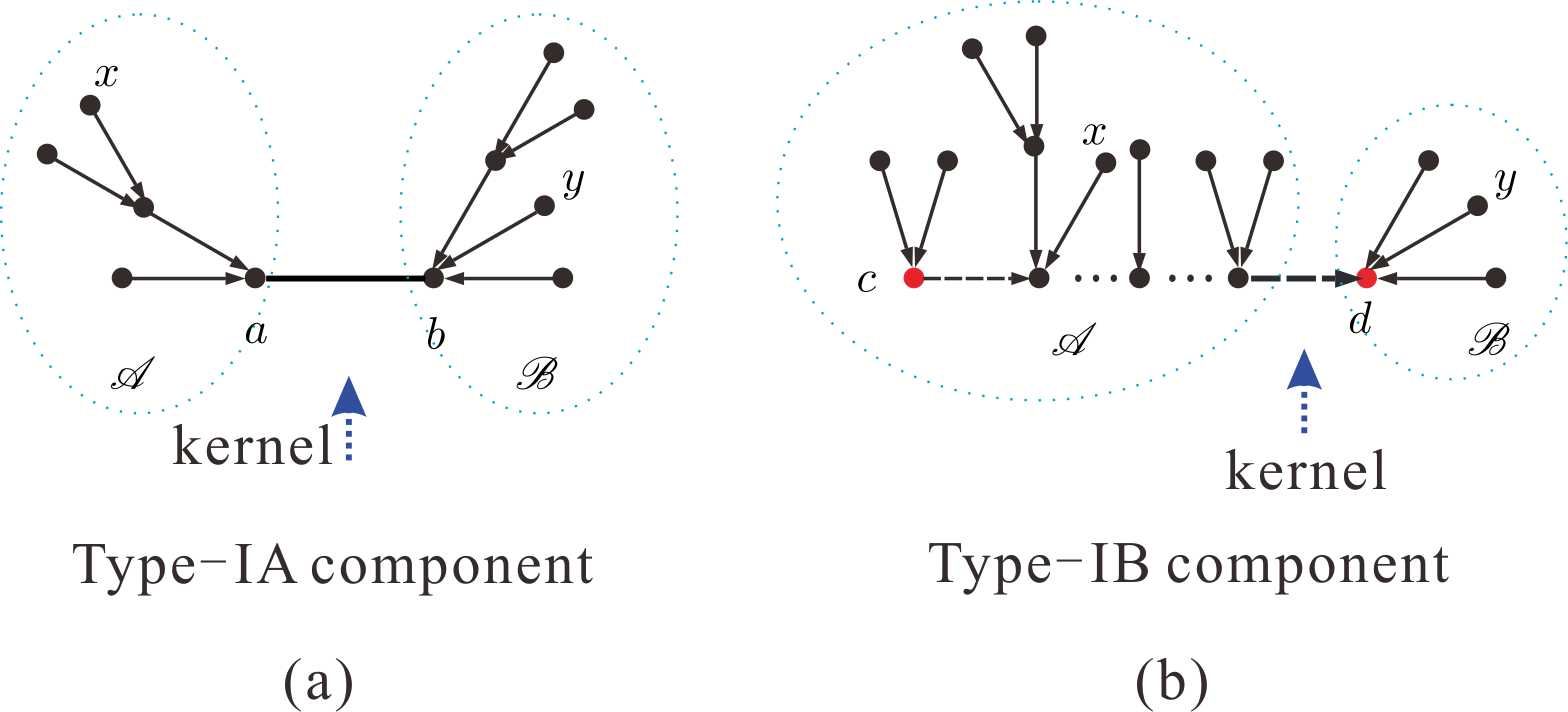}
	\caption{Graph (a) is a type-IA component, while graph (b) is a type-IB component. The kernel of (a) is defined by the type-1 line between nodes $a$ and $b$. The kernel of (b) is defined by the type-4 line which is connected to the (unfixed) end node $d$. The nodes $x$ and $y$ in each graph are supposed to be the highest-weight nodes in the regions $\mathscr{A}$ and $\mathscr{B}$, respectively. If the weight of $x$ is higher than that of $y$, $\mathscr{A}$ and $\mathscr{B}$ are correspondingly the top and the bottom sides. Contrarily, if the weight of $y$ is higher than that of $x$, $\mathscr{A}$ becomes the bottom side while $\mathscr{B}$ the top.  }\label{Fig:type-1Components}
\end{figure}
\begin{figure}
	\centering
	\includegraphics[width=0.8\textwidth]{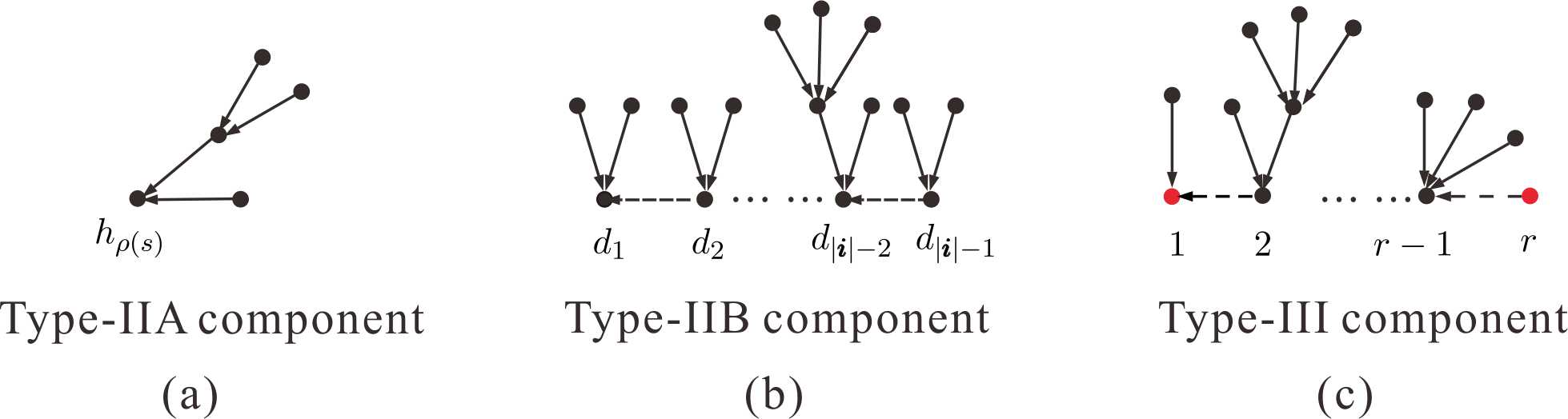}
	\caption{The component involving the highest-weight element $\mathcal{H}_{\rho(l)}$ is defined as {(a).} the type-IIA component if $\mathcal{H}_{\rho(l)}$ is a graviton $h_{\rho(s)}$, {(b).} the type-IIB component if $\mathcal{H}_{\rho(l)}$ is a gluon trace $\pmb{t}\to \{d_1,\dots,d_{|\pmb{t}|-1}\}$. The component containing the trace $\pmb{1}=\{1,2,\dots,r-1,r\}$, as shown by the graph (c), is defined as the type-III component.}\label{Fig:type-2-3Components}
\end{figure}
\subsection{Skeletons and components}\label{Sec:SkeletonsComponents}
When all type-3 lines (i.e. $k\cdot k$ factors) are removed, a graph $\mathcal{G}$ becomes a skeleton $\mathcal{G}'$. To analyze possible structures of components which are maximally connected subgraphs of $\mathcal{G}'$, we should look into the inner structure of a chain via expanding \eqref{Eq:Chain2} by \eqref{Eq:EFY1}, \eqref{Eq:EFY2} and \eqref{Eq:SplittingTraces}. According to whether the starting node is the highest-weight element in the reference order $\mathsf{R}$ (defined in \eqref{Eq:ReferenceOrder}) or not, we carry out the discussion as follows:
\begin{itemize}
\item {\bf(i).} If the starting node of a chain is not the highest-weight element $\mathcal{H}_{\rho(l)}$, a graph corresponding to this chain can only have the general pattern \figref{Fig:ChainStructure1} (a). Now we comment on crucial features of \figref{Fig:ChainStructure1} (a): {(1).} Distinct sectors are separated by type-3 lines. The sector containing the starting (ending) node is called the \emph{starting (ending) sector}. Other sectors between the starting and ending sectors are mentioned as \emph{internal sectors}. {(2).} The possible structures inside the  boxes in \figref{Fig:ChainStructure1} (a) are given by \figref{Fig:ChainStructure1} (d) (where two gravitons are connected by a type-1 line) and \figref{Fig:ChainStructure1} (e), (f) (which involve a gluon trace expressed by standard basis).
    {(3).} If a chain \figref{Fig:ChainStructure1} (a) contains only one sector, the sector must be the ending sector. {(4)}. If the starting element of a chain is a gluon trace, there must be no graviton on the left hand side of the box in the starting sector, as shown by \figref{Fig:ChainStructure1} (f).
    {(5)}. Other structures of a chain are forbidden because the substructures in \figref{Fig:ChainStructure1} (c) are not allowed by   (\eqref{Eq:Chain2}) (in other words, $F_{h}^{\mu\nu}\equiv k_h^{\mu}\epsilon_h^{\nu}-\epsilon_h^{\mu}k_h^{\mu}$ for a graviton $h$  involves neither $k^{\mu}k^{\nu}$ nor $\epsilon_h^{\mu}\epsilon_h^{\nu}$).


\item {\bf(ii).} If the starting node of a chain is the highest-weight element $\mathcal{H}_{\rho(l)}$, its corresponding graph must have the general pattern \figref{Fig:ChainStructure1} (b), where the highest-weight element (graviton or trace) is already replaced according to \figref{Fig:Replacement} (a) (for a graviton) or \figref{Fig:Replacement} (b) and (c) (for a gluon trace). An important feature is \emph{the  chain \figref{Fig:ChainStructure1} (b)  has at least two sectors},  which follows from the fact that structures in \figref{Fig:ChainStructure1} (c) are forbidden.
\end{itemize}

Having the above discussions, sectors of chains can be easily classified as \figref{Fig:ChainStructure3}. In a full graph $\mathcal{G}$, nodes of any sector may play as the ending nodes of other chains. When all type-3 lines are removed, each sector in a skeleton thus can be attached by the type-3 sectors \figref{Fig:ChainStructure3} (c) (or equivalently ending sectors) of other chains. Consequently, components in a skeleton can be classified by the following way.
\begin{itemize}
\item{\bf Type-I component:} \emph{A component consisting of a type-I sector \figref{Fig:ChainStructure3} (a) and possible type-III sectors \figref{Fig:ChainStructure3} (c) whose arrow lines point towards the type-I sector}~~~Type-I components can further be classified according to different structures inside the box of \figref{Fig:ChainStructure3} (a): If the box contains the structure \figref{Fig:ChainStructure1} (d), as shown by \figref{Fig:type-1Components} (a), this component is called a {\bf type-IA component}. Else, if the structure in the box is given by \figref{Fig:ChainStructure1} (e) or (f), as shown by \figref{Fig:type-1Components} (b), the component is called a {\bf type-IB component}. We define the {\bf kernel} of a type-I component by (i). the type-1 line of a type-IA component (see \figref{Fig:type-1Components} (a)), (ii). the type-4 line that is attached to the unfixed end node of the trace (in standard basis) inside a type-IB component (see \figref{Fig:type-1Components} (b)). For a given reference order, any type-IA and -IB component is divided into two parts by the kernel: the part involving the highest-weight node (although a trace is considered as a single object in the reference order, the fixed node $c$ is always considered as the highest-weight node of this trace and it carries the weight of the full trace in the reference order) of this component is called the {\bf{top side}}, while the opposite part is called the {\bf{bottom side}}.

\item {\bf Type-II component:} \emph{A component consisting of a type-II sector \figref{Fig:ChainStructure3} (a) and possible type-III sectors whose arrows point towards the type-II sector}~~~ If the structure in the box of the type-II sector is \figref{Fig:ChainStructure1} (g) (i.e. the highest-weight element is a graviton), this component is called a \textbf{type-IIA component} (see \figref{Fig:type-2-3Components} (a)). If the structure in the box is a gluon trace \figref{Fig:ChainStructure1} (h), the component is called a \textbf{type-IIB component} (see \figref{Fig:type-2-3Components} (b)).

\item {\bf Type-III component:} \emph{A component consisting of the trace $\pmb{1}$ and possible type-III sectors with arrows pointing towards the trace $\pmb{1}$  (see \figref{Fig:type-2-3Components} (c))}
\end{itemize}

Since the chain \figref{Fig:ChainStructure1} (b) that is lead by the highest weight node has at least two sectors, a skeleton must at least contain two components: the type-II and the type-III components. In general, Type-I components may also be involved in a skeleton. All those graphs corresponding to a given skeleton are reproduced by connecting type-3 lines between components of a skeleton $\mathcal{G}'$ in an appropriate way.

\subsubsection*{Multi-trace from single-trace}
It is worth pointing out that the refined graphic rule given in \secref{Sec:RefinedGraphicRule} can be obtained from the rule for identities induced from single-trace amplitudes (which was presented in \cite{Hou:2018bwm}) by an appropriate replacement. Particularly, we consider a single-trace amplitude $A(1,2,\dots,r\Vert \mathsf{H})$ where we have $s+m-1$ gravitons $\mathsf{H}=\{h_1,\dots,h_{s+m-1}\}$. When we replace $m-1$ gravitons, e.g., $h_{s+1}, h_{s+2},\dots,h_{s+m-1}$ by $m-1$ gluon traces $\pmb{2},\dots,\pmb{m}$, we get the multi-trace amplitude $A(1,2,\dots,r|\pmb{2}|\dots|\pmb{m}\Vert \mathsf{H})$ with $\mathsf{H}=\{h_1,\dots,h_{s}\}$. Such replacement is reflected in the refined graphic rule via replacing the reference order and chains with only gravitons by those with gravitons and/or gluon traces. The corresponding replacement for coefficients are given by
\bea
\epsilon\cdot F\cdot F\cdot \ldots\cdot F\cdot k \xrightarrow{h_{s+1}, \dots,h_{s+m-1}\to \pmb{2},\dots,\pmb{m}} \mathcal{E}\cdot \mathbb{F}\cdot \mathbb{F}\cdot...\cdot \mathbb{F}\cdot k,\Label{Eq:Replacement1}
\eea
where $\epsilon$, $F$ are the half polarizations and strength tensors of gravitons, while  $\mathcal{E}$ and  $\mathbb{F}$ are generalized polarizations and generalized strength tensors for gravitons and/or gluon traces (see \eqref{Eq:EFY1} and \eqref{Eq:EFY2}). Graphically, this replacement is given by
\bea
\text{\figref{Fig:RefinedGraviton} (a)}&\to& \text{\figref{Fig:RefinedTrace} (a)}, (\text{if $h_i$ is replaced by $\pmb{t}_i$})\nn
\text{\figref{Fig:RefinedGraviton} (b)}&\to& \text{\figref{Fig:RefinedTrace} (b)}, (\text{if $h_i$ is replaced by $\pmb{t}_i$}).
\eea
For identities induced from a single-trace amplitude, the highest-weight element (graviton) is further given by $k^{\mu}$ (see \figref{Fig:Replacement} (a)). In the corresponding multi-trace case, the highest-weight element can either be a graviton (if it is not replaced by a trace) or be replaced by a gluon trace. The latter is described via replacing the node (the highest-weight graviton for identity induced from single-trace amplitude) by the RHS of \figref{Fig:Replacement} (c).

Now we look into the inner structure of a chain by further expanding internal gravitons on the LHS of \eqref{Eq:Replacement1} according to  $F^{\mu\nu}=k^{\mu}\epsilon^{\nu}-\epsilon^{\mu}k^{\nu}\equiv F_{(+)}^{\mu\nu}-F_{(-)}^{\mu\nu}$. On the RHS, internal gravitons are also expanded by $F_{(+)}^{\mu\nu}-F_{(-)}^{\mu\nu}$, while internal gluon traces are expanded according to the relation \eqref{Eq:SplittingTraces}.
Although this expansion of gluon trace does not affect the tensor $\mathbb{F}^{\mu\nu}$ for a given $a_i$ and $b_i$ (see \eqref{Eq:EFY2}), it splits the graphs corresponding to the trace into the standard basis. Hence, the replacement $h_i\to \pmb{i}$ (for an internal graviton $h_i$) is achieved graphically through replacing the first (second) graph in \figref{Fig:RefinedGraviton} (b) by the first (second) graph on the RHS of  \eqref{Eq:SplittingTraces} (for a given $\{a_i,b_i\}$). It follows that the general chain structures \figref{Fig:ChainStructure1} (a) and (b) are obtained from those chain structures for the single-trace case \cite{Hou:2018bwm} (where only type-IA, type-IIA and type-III sectors are allowed) by incorporating more types of sectors: type-IB and type-IIB sectors which reflect structures of gluon traces.  Consequently, the full classification of components in multi-trace cases can be given by enlarging the families of the type-I and the type-II components that were defined in \cite{Hou:2018bwm}:
\bea
    \text{Type-I components}&\to&\text{Type-IA or Type-IB components},\nn
    \text{Type-II components}&\to&\text{Type-IIA or Type-IIB components},\Label{Eq:Enlargement}
    \eea
   where the type-IIA and type-IIB components are respectively the highest-weight components of the identities (\ref{Eq:InducedID1}) and  (\ref{Eq:InducedID2}). All the above discussions allow us to borrow some crucial conclusions from the single-trace case \cite{Hou:2018bwm}:
\begin{itemize}
\item {\bf (i)} When keeping track of chains in the single-trace case \cite{Hou:2018bwm}, one can build all possible physical graphs corresponding to a skeleton  by connecting type-3 lines between components properly (see appendix D in \cite{Hou:2018bwm}). This construction can be immediately generalized to multi-trace cases by the enlargement (\ref{Eq:Enlargement}).

\item {\bf (ii)} As proved in \cite{Hou:2018bwm} (see sections 6.1, 6.2 and appendix D of \cite{Hou:2018bwm}), all the physical graphs in the single-trace case, which are corresponding to a given skeleton and are constructed by the above step, can be reproduced by (1). constructing the final upper and lower blocks, (2). connecting the final upper and lower blocks via a type-3 line appropriately. In multi-trace cases, we just follow the same construction rule but enlarging the type-I and type-II classes of components according to (\ref{Eq:Enlargement}).

\item {\bf (iii)} As pointed in section 6.3 of \cite{Hou:2018bwm}, in the single-trace case, the sum over all physical graphs which are produced by (ii) can be further written as the sum of all physical and spurious graphs (those graphs which are not directly constructed from the refined graphic rule). The latter all cancel out after summation. Again, the spurious graphs for single-trace induced identities can  be straightforwardly extended to multi-trace cases by the help of (\ref{Eq:Enlargement}) and they all cancel out (we have seen this cancellation by the examples in \secref{Sec:Examples}).
\end{itemize}
In the coming two subsections, we display the construction rule of the final upper and lower blocks as well as the construction of physical and spurious graphs without a proof. In fact, all the proofs follow from discussions parallel with those in the single-trace case \cite{Hou:2018bwm}.

\subsection{The final upper and lower blocks}\label{Sec:FinalBlocks}

Now we  provide the general rule for constructing all possible configurations of the final upper and lower blocks  corresponding to a given skeleton $\mathcal{G}'$:
\begin{itemize}
\item {\bf{Step-1}} For any skeleton $\mathcal{G}'$, we define \emph{the reference order $\mathsf{R}_{\mathscr{C}}$ of all type-I components (including type-IA and type-IB components)} by the relative order of the highest-weight nodes therein. In other words, the weight (i.e. the position in $\mathsf{R}_{\mathscr{C}}$) of a component inherits from its highest-weight node. For example, suppose there are three type-I components (IA and/or IB) $\mathscr{C}_1$, $\mathscr{C}_2$, and $\mathscr{C}_3$ with the corresponding highest-weight nodes (graviton or a gluon) $a_1$, $a_2$ and $a_3$.
    If the weights $W_{a_i}$  have the relation $W_{a_2}<W_{a_1}<W_{a_3}$, the reference order of these components is then given by the ordered set $\mathsf{R}_{\mathscr{C}}=\{\mathscr{C}_2,\mathscr{C}_1,\mathscr{C}_3\}$. We further define the upper block $\mathscr{U}$ and lower block $\mathscr{L}$ as the components respectively containing the highest-weight element (graviton or trace) and the trace $\pmb{1}$. At the beginning, the upper and the lower blocks are nothing but the type-II and the type-III components.

\item {\bf{Step-2}} Supposing the reference order of components is $\mathsf{R}_{\mathscr{C}}=\left\{\mathscr{C}_1,\mathscr{C}_2,\dots,\mathscr{C}_N\right\}$, pick out the highest-weight component $\mathscr{C}_N$ as well as arbitrary components $\mathscr{C}_{a_1}$, $\mathscr{C}_{a_2}$, ..., $\mathscr{C}_{a_i}$ (the relative order of these components is not necessary the same relative order in $\mathsf{R}_{\mathscr{C}}$).
    Construct a chain of components towards either the upper block or the lower block as follows
    \bea
    \mathbb{CH}=\left[(\mathscr{C}_N)_t,(\mathscr{C}_N)_b\leftrightarrow(\mathscr{C}_{a_i})_{t\,(\text{or}\,b)},(\mathscr{C}_{a_i})_{b\,(\text{or}\,t)}\leftrightarrow\cdots\leftrightarrow(\mathscr{C}_{a_1})_{t\,(\text{or}\,b)},(\mathscr{C}_{a_1})_{b\,(\text{or}\,t)}\leftrightarrow\mathscr{U}\,\text {or}\,\mathscr{L}\setminus\{r\}\right].\nn
    \eea
 Here the subscripts $t$ and $b$ respectively denote the top and bottom sides of a type-I component, which are separated by a comma\footnote{Notations here are slightly different from those in \cite{Hou:2018bwm}.}. The double arrow line `$\leftrightarrow$' between two components stands for the type-3 line (i.e. $k\cdot k$), which connects any two nodes belonging to the corresponding regions. For example, if the chain of components has the form $\left[(\mathscr{C}_N)_t,(\mathscr{C}_N)_b\leftrightarrow(\mathscr{C}_{a_i})_{t},(\mathscr{C}_{a_i})_{b}\leftrightarrow\cdots\right]$, the two ends $x$ and $y$ of the type-3 line between the components $\mathscr{C}_N$ and $\mathscr{C}_{a_i}$ must belong to $(\mathscr{C}_N)_b$ and $(\mathscr{C}_{a_i})_{t}$ respectively. After this step, we redefine the reference order of components as well as the upper and lower blocks by:
 \bea
 \mathsf{R}_{\mathscr{C}}&\to&\mathsf{R}'_{\mathscr{C}}=\mathsf{R}_{\mathscr{C}}\setminus\{\mathscr{C}_N,\mathscr{C}_{a_i},\dots,\mathscr{C}_{a_1}\}=\left\{\mathscr{C}'_1,\mathscr{C}'_2,\dots,\mathscr{C}'_N\right\},\nn
 \mathscr{U}&\to&\mathscr{U}'=\mathscr{U}\cup\{\mathscr{C}_N,\mathscr{C}_{a_i},\dots,\mathscr{C}_{a_1}\},~~~~\mathscr{L}\to\mathscr{L}'=\mathscr{L}~~\text{(if $\mathbb{CH}$ was attached to $\mathscr{U}$)}\nn
 \mathscr{L}&\to&\mathscr{L}'=\mathscr{L}\cup\{\mathscr{C}_N,\mathscr{C}_{a_i},\dots,\mathscr{C}_{a_1}\},~~~~\mathscr{U}\to\mathscr{U}'=\mathscr{U}~~\text{(if $\mathbb{CH}$ was attached to $\mathscr{L}$)}.
 \eea

\item {\bf{Step-3}} Repeating step-2 with the new defined $\mathsf{R}_{\mathscr{C}}$, $\mathscr{U}$ and $\mathscr{L}$ iteratively until the ordered set $\mathsf{R}$ becomes empty, we get a graph with only two mutually disjoint subgraphs: the final upper and lower blocks $\mathscr{U}$ and $\mathscr{L}$.
\end{itemize}
\begin{figure}
	\centering
	\includegraphics[width=1\textwidth]{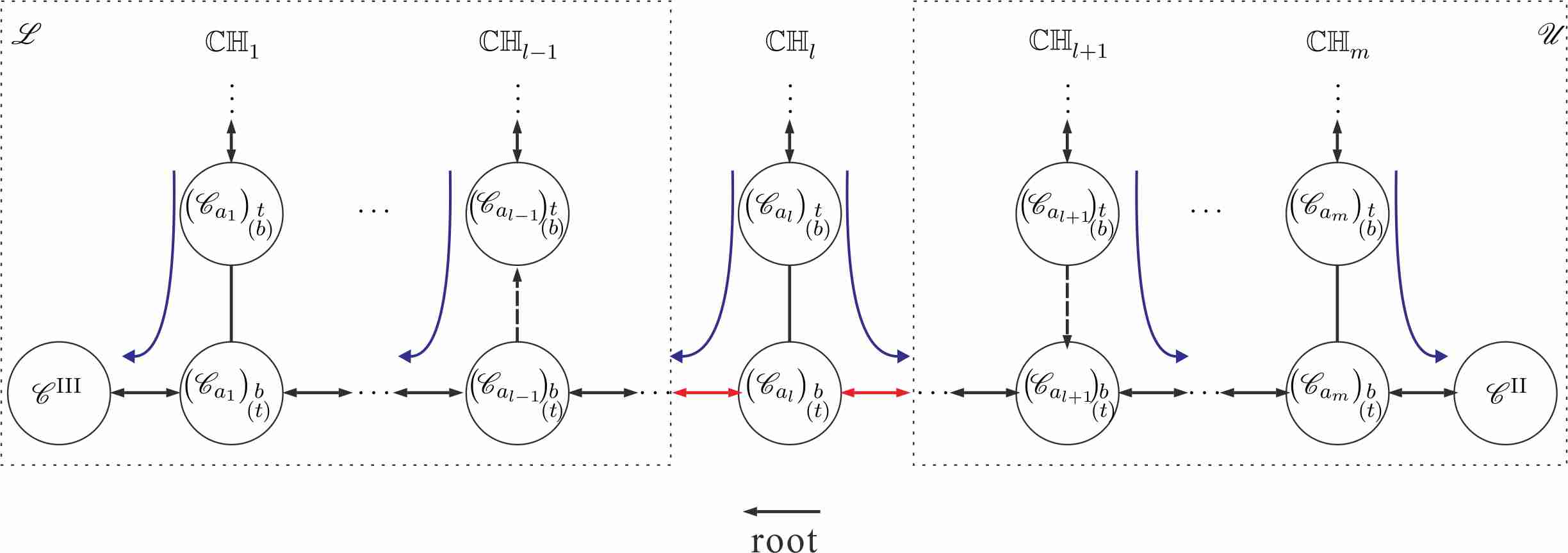}
	\caption{ A typical spurious graph where the path starting from the highest-weight element and ending at the root $1$ passes through some single sides of type-I (IA or IB) components $\mathscr{C}_{a_1}, \mathscr{C}_{a_2}, \dots, \mathscr{C}_{a_m}$ (which are called spurious components).}\label{Fig:Spurious}

\end{figure}
All possible configurations of the final upper and lower blocks are produced by the above steps.
The final upper and lower blocks for the examples in \secref{Sec:Examples} are precisely reproduced by this rule (see \figref{Fig:FinalUL1} (a), (b) for example-1 and \figref{Fig:SkeletonEG} (b), (c) for example-2).

\subsection{Physical and spurious graphs}

For a given configuration of the final upper and lower blocks $\mathscr{U}$ and $\mathscr{L}$ which are constructed previously, a fully connected graph $\mathcal{G}$ in \eqref{Eq:InducedIDUnified3} is produced by connecting arbitrary two nodes $x\in\mathscr{U}$ and $y\in\mathscr{L}\setminus\{r\}$ via a type-3 line. As pointed in \secref{Sec:Examples}, such a graph can be either a physical graph or a spurious one. We have already  stated that physical and spurious graphs can be obtained from those in single-trace case \cite{Hou:2018bwm} by the enlargement (\ref{Eq:Enlargement}). As a result, a spurious graph has the structure \figref{Fig:Spurious}, where the chain starting form the highest-weight element (graviton for the identity (\ref{Eq:InducedID1}) and gluon for the identity (\ref{Eq:InducedID2})) and ending at the root $1$ passes through single sides of some type-IA and/or type-IB components $\mathscr{C}_{a_1}, \mathscr{C}_{a_2}, \dots, \mathscr{C}_{a_m}$ which respectively belong to the chains of components $\mathbb{CH}_1,\dots, \mathbb{CH}_m$.

In order to display more details of spurious graphs, we define the weight $W_i$ of a chain $\mathbb{CH}_{a_i}$
by the weight of  the starting component (equivalently the highest-weight component) of $\mathbb{CH}_{a_i}$. As pointed in \cite{Hou:2018bwm}, if the lowest-weight chain among $\mathbb{CH}_1,\dots, \mathbb{CH}_m$ in \figref{Fig:Spurious} is $\mathbb{CH}_l$, we must have $W_1>\dots>W_{l-1}>W_l$ and $W_{l}<W_{l+1}<\dots<W_m$. Following a discussion which is parallel with that in \cite{Hou:2018bwm}, we conclude that all the chains  $\mathbb{CH}_1,\dots, \mathbb{CH}_{l-1}$ (and structures attached to them) belong to the final lower block $\mathscr{L}$, while $\mathbb{CH}_{l+1},\dots, \mathbb{CH}_{m}$ (and structures attached to them) belong to the final upper block $\mathscr{U}$. Only the lowest-weight chain $\mathbb{CH}_l$ (among $\mathbb{CH}_{l+1},\dots, \mathbb{CH}_{m}$) can live in either $\mathscr{L}$ or $\mathscr{U}$. Correspondingly, the type-3 line (colored by red in \figref{Fig:Spurious}) on either the LHS or the RHS of  $\mathscr{C}_{a_l}$ is considered as the one between the final upper and lower blocks. Thus a given spurious graph is corresponding to two distinct configurations of $\mathscr{U}$ and $\mathscr{L}$. In other words, all spurious graphs must appear in pairs! This fact allows us to associate a pair of spurious graphs with opposite signs so that all spurious graphs cancel out.



By the help of the above discussion, we now determine the sign $(-)^{\mathcal{G}}$ in \eqref{Eq:InducedIDUnified3} for a (physical or spurious) graph $\mathcal{G}$:
\begin{itemize}
     \item {\bf(i).} As proposed in \cite{Hou:2018bwm}, a graph is accompanied by a sign $(-1)^{S(\mathscr{U}_x)}$, where $S(\mathscr{U}_x)$ is the number of spurious components living in the final upper block $\mathscr{U}$ for a given $x\in\mathscr{U}$ in \eqref{Eq:InducedIDUnified3}. For the typical spurious graph \figref{Fig:Spurious}, $S(\mathscr{U}_x)=(-1)^{m-l+1}$, if $\mathscr{C}_{a_l}$ belongs to the final upper block $\mathscr{U}$, while $S(\mathscr{U}_x)=(-1)^{m-l}$, if $\mathscr{C}_{a_l}$ belongs to the final lower block $\mathscr{L}$. Hence spurious graphs cancel in pairs precisely.

    \item {\bf(ii).} Another sign which should be taken into account is introduced by the relation \eqref{Eq:SplittingTraces}. Particularly, if the arrow of the kernel in a type-IB component (see \figref{Fig:type-1Components} (b)) is pointing away from the root (i.e. the second term of \eqref{Eq:SplittingTraces}), this component should be dressed by an extra minus. Since the choice of $y\in\mathscr{L}\setminus\{r\}$ in \eqref{Eq:InducedIDUnified3} does not affect the direction of arrows in the final lower block $\mathscr{L}$, the number of such type-IB components only depends on $\mathscr{U}$, $\mathscr{L}$ and the choice of $x\in \mathscr{U}$ in \eqref{Eq:InducedIDUnified3}. We use $Tr(\mathscr{U}_x,\mathscr{L})$ to denote this number, the resulting sign is then written as $(-1)^{Tr(\mathscr{U}_x,\mathscr{L})}$.

    \item {\bf(iii).} For any graph, the total number $\mathcal{N}(\mathscr{U}_x)+\mathcal{N}(\mathscr{L})+1$ of arrows pointing away from the root induces the third sign $(-1)^{\mathcal{N}(\mathscr{U}_x)+\mathcal{N}(\mathscr{L})+1}$ (as required by the refined graphic rule), where $\mathcal{N}(\mathscr{U}_x)$ and $\mathcal{N}(\mathscr{L})$ count the corresponding numbers in $\mathscr{U}$ (for $x\in \mathscr{U}$) and $\mathscr{L}$. The extra minus is caused by the type-3 line between $\mathscr{U}$ and $\mathscr{L}$.
   \end{itemize}
To sum up, the sign for any (physical or spurious) graph is given by
\bea
(-)^{\mathcal{G}}=(-1)^{S(\mathscr{U}_x)+Tr(\mathscr{U}_x,\mathscr{L})+\mathcal{N}(\mathscr{U}_x)+\mathcal{N}(\mathscr{L})+1}.\Label{Eq:Sign}
\eea
For a physical graph $\mathcal{G}$, there is no spurious component, thus $S(\mathscr{U}_x)=0$. For identities induced from single-trace amplitudes, $Tr(\mathscr{U}_x,\mathscr{L})$ vanishes and the sign (\ref{Eq:Sign}) turns into the one given in \cite{Hou:2018bwm}.

\subsection{Expanding induced identities in terms of BCJ relations}\label{Sec:InducedBCJ}
\begin{figure}
	\centering
	\includegraphics[width=0.46\textwidth]{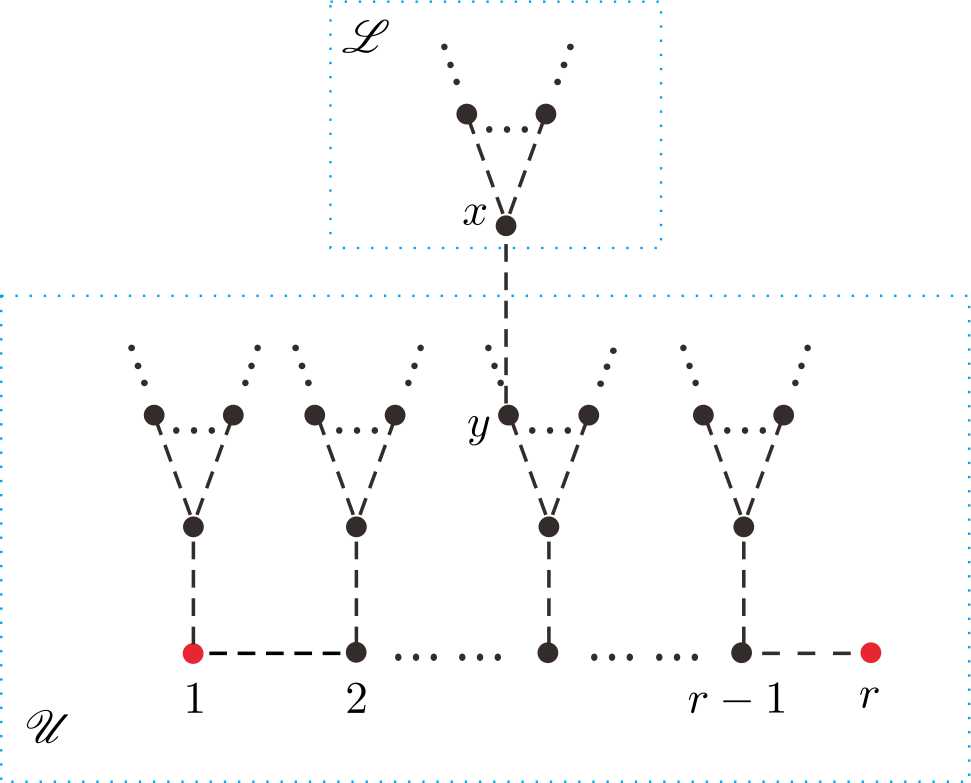}
	\caption{ Relative positions of nodes in a graph $\mathcal{G}$ which is determined by $\mathscr{U}$, $\mathscr{L}$ and $x\in \mathscr{U}$, $y\in \mathscr{L}\setminus\{r\}$ can be characterized by dashed lines with no arrow.  }\label{Fig:GeneralPatternG}
\end{figure}

We are now ready to show the expression in the square brackets in \eqref{Eq:InducedIDUnified3}, i.e.
\bea
I[\mathscr{U},\mathscr{L}]\equiv\Sl_{\substack{x\in\mathscr{U}\\y\in\mathscr{L}\setminus\{r\}}}\,\Sl_{\pmb{\sigma}^{\mathcal{G}}}(-)^{\mathcal{G}}\,(k_x\cdot k_y)\,A\big(1,\pmb{\sigma}^{\mathcal{G}},r\big)~\Label{Eq:InducedIDUnified4}
\eea
for a given configuration of the final upper and lower blocks $\mathscr{U}$, $\mathscr{L}$ is the LHS of the graph-based BCJ relation (\ref{Eq:Graph-based-BCJ}) (hence a combination of traditional BCJ relations (\ref{Eq:BCJRelation})). Here a graph $\mathcal{G}$ in the above expression is a physical or a spurious graph which is constructed by connecting two nodes $x\in \mathscr{U}$, $y\in\mathscr{L}\setminus\{r\}$ via a type-3 line. Our discussion is carried out by the follow steps:

\begin{itemize}
\item {\bf(i).} According to the refined graphic rule, permutations $\pmb{\sigma}^{\mathcal{G}}$ in \eqref{Eq:InducedIDUnified4} are independent of the line styles in $\mathcal{G}$. Therefore, all lines in ${\mathcal{G}}$ can be replaced by dashed lines (with no arrow) which only characterize the relative positions of nodes in ${\mathcal{G}}$, as shown by \figref{Fig:GeneralPatternG}. Apparently, for a given configuration of the final upper and lower blocks $\mathscr{U}$, $\mathscr{L}$ and a given choice of $x\in \mathscr{L}$,
    the corresponding permutations $\pmb{\sigma}^{\mathcal{G}}$ satisfy
\bea
\pmb{\sigma}^{\mathcal{G}}\in \Bigl[\pmb{\zeta}\shuffle\pmb{\gamma}\Bigr]\Big|_{y\prec x}~~~~\big(\text{for~} \pmb{\zeta}\in{\mathscr{U}}\big|_{x} \text{~and~} \pmb{\gamma}\in\bigl(\mathscr{L}\big|_1\bigr)\setminus\{1,r\}\big),\Label{Eq:GenPermutationPattern}
\eea
where ${\mathscr{U}}\big|_{x}$ and $\mathscr{L}\big|_1$ denote the permutations established by $\mathscr{U}$ and $\mathscr{L}$ when $x\in \mathscr{U}$ and the root $1\in \mathscr{L}$ are the leftmost elements respectively.
Since $1$ and $r$ are fixed as the first and the last elements in \eqref{Eq:InducedIDUnified4}, they should be excluded from $\pmb{\sigma}^{\mathcal{G}}$. The $y\prec x$ means $y$ situates before $x$ in the permutation, i.e. $(\sigma^{\mathcal{G}})^{-1}(y)<(\sigma^{\mathcal{G}})^{-1}(x)$. Noting that the choice of $y$ is independent of the relative orders  $\pmb{\zeta}$ and $\pmb{\gamma}$, we rewrite the summations in \eqref{Eq:InducedIDUnified4} as follows
\bea
\Sl_{\substack{x\in\mathscr{U}\\y\in\mathscr{L}\setminus\{r\}}}\,\Sl_{\pmb{\sigma}^{\mathcal{G}}}\to \Sl_{x\in \mathscr{U}}\,\Sl_{\pmb{\zeta}\in\mathscr{U}|_{x}}\,\Sl_{\pmb{\gamma}\in\mathscr{L}|_{1}\setminus\{1,r\}}\,\Sl_{y\in \mathscr{L}\setminus\{r\}}\,\Sl_{\pmb{\sigma}\in [\pmb{\zeta}\shuffle\pmb{\gamma}]|_{y\prec x}}.\Label{Eq:RearrangeSummation}
\eea

\item {\bf(ii).} For a given permutation $\pmb{\sigma}\in \pmb{\zeta}\shuffle\pmb{\gamma}$, one can collect together the coefficients $k_x\cdot k_y$ in \eqref{Eq:GenPermutationPattern} with different choices of $y\in \mathscr{L}\setminus\{r\}$. Specifically, only those $y$ satisfying $y\prec x$ in $\pmb{\sigma}$ have nonzero contributions and the sign (\ref{Eq:Sign}) is totally independent of $y$. Hence, all the $k_x\cdot k_y$ factors for a given $\pmb{\sigma}$ are collected as  $k_x\cdot Y_x(\pmb{\sigma})$ where $Y^{\mu}_x(\pmb{\sigma})\equiv\sum_{y\prec x}k_y^{\mu}$ (the gluon $1$ is always included as the leftmost $y$ in this summation). Meanwhile, the last two summations in \eqref{Eq:RearrangeSummation} turns into
\bea
\Sl_{y\in \mathscr{L}\setminus\{r\}}\,\Sl_{\pmb{\sigma}\in [\pmb{\zeta}\shuffle\pmb{\gamma}]|_{y\prec x}}\to \Sl_{\pmb{\sigma}\in \pmb{\zeta}\shuffle\pmb{\gamma}}.\Label{Eq:RearrangeSummation1}
\eea

\item {\bf(iii).} As illustrated by \appref{sec:Sign}, the sign $(-)^{\mathcal{G}}$ in \eqref{Eq:InducedIDUnified4} has the following pattern. Two graphs with $x=x_1$ and $x=x_2$ where $x_1,x_2\in \mathscr{U}$ are adjacent to each other must be associated with opposite signs.
Hence one can extract the sign \eqref{Eq:Sign} for a graph  with $x=x_0\in \mathscr{U}$ as an overall factor and then the sign for an arbitrary choice of node $x$ is given by
\bea
(-1)^{S(\mathscr{U}_{x_0})+Tr(\mathscr{U}_{x_0},\mathscr{L})+\mathcal{N}(\mathscr{U}_{x_0})+\mathcal{N}(\mathscr{L})+1}f^x\equiv (-)^{\mathcal{G}({x_0})}f^x,
\eea
where $f^x$ is defined by {(i).} $f^{x_0}=1$, {(ii).} $f^{x_1}=-f^{x_2}$ for  two adjacent nodes $x_1$ and $x_2$ ($x_1,x_2\in \mathscr{U}$).
\end{itemize}
When all the above are taken into account, \eqref{Eq:InducedIDUnified4} is finally expressed by
\bea
I[\mathscr{U},\mathscr{L}]=(-)^{\mathcal{G}({x_0})}\Sl_{\pmb{\gamma}\in\mathscr{L}|_1\setminus\{1,r\}}\biggl[\Sl_{x\in \mathscr{U}}f^x\Sl_{\pmb{\zeta}\in\mathscr{U}|_x}\Sl_{\pmb{\sigma}\in \pmb{\zeta}\shuffle\pmb{\gamma}}(k_x\cdot Y_{x}(\pmb{\sigma}))A\Bigl(1,\pmb{\sigma},r\Bigr)\biggr],
\Label{Eq:InducedID6}
\eea
where the summation over $\pmb{\gamma}$ was extracted out because it is independent of the choice of $x\in \mathscr{U}$. The expression in the square brackets in \eqref{Eq:InducedID6} is nothing but the LHS of the graph-based BCJ relation (\ref{Eq:Graph-based-BCJ}).

\section{Conclusions}\label{sec:Conclusions}
In this paper, we provided the refined graphic rule for expanding tree level multi-trace EYM amplitudes in terms of color-ordered YM amplitudes. When the gauge invariance condition of a graviton and the cyclic symmetry of a gluon trace were imposed, this expansion induced two identities (\ref{Eq:InducedID1}) and (\ref{Eq:InducedID2}) respectively. By extending the analysis for the single-trace case \cite{Hou:2018bwm} to an arbitrary multi-trace induced identity, we demonstrated that \eqref{Eq:InducedID1} and \eqref{Eq:InducedID2} can finally be expressed as a combination of graph-based BCJ relations (thus traditional BCJ relations).


There are several related topics that deserve further study: (i). First, how to understand the induced identities from the view of string theory?
String theory studies of the expansions of EYM amplitudes have been established in \cite{Stieberger:2016lng,Schlotterer:2016cxa,He:2018pol,He:2019drm}, while BCJ relations have also been proven in string theory \cite{BjerrumBohr:2009rd,Stieberger:2009hq}. Hence it is reasonable to expect a string-theory approach to both induced identities and graph-based BCJ relations. (ii). Second, it is worth investigating the induced identities in various theories systematically. In \cite{Zhou:2019mbe}, a unified web of expansions of amplitudes was founded with the help of the unifying relation \cite{Cheung:2017ems}, which inspires that the induced identities may exist in many other theories.  (iii). Third,  the YM expansion of EYM amplitudes, which have been used in this paper, is in KK basis \cite{Kleiss:1988ne}. As pointed in \cite{Feng:2019tvb}, this expansion can be extended to BCJ basis \cite{Bern:2008qj}. We expect that the refined graphic rule can also be generalized to expansions in BCJ basis \cite{Bern:2008qj}.  (iv). Last but not least, a kinematic algebra for constructing BCJ numerators in the MHV sector was proposed \cite{Chen:2019ywi}. It seems that distinct sectors of numerators are corresponding to graphs with different numbers of type-IA kernels. Thus, one may provide a general rule for constructing all sectors of BCJ numerators, with the help of refined graphic rule.

\section*{Acknowledgments}
The authors are grateful to Chih-Hao Fu, Song He, Xiaodi Li, Hui Luo, Gang Yang, Yihong Wang and Yong Zhang for helpful discussions or/and valuable comments.
This work is supported by NSFC under Grant Nos. 11875206, 11847309, Jiangsu Ministry of Science and Technology under contract
BK20170410 as well as the ``Fundamental Research Funds for the Central Universities".

\appendix

\section{Recursive expansions and BCJ relations}\label{sec:Review}
In this section, we review the recursive expansions of multi-trace EYM amplitudes and the BCJ relations for YM amplitudes.

\subsection{Recursive expansions of multi-trace EYM amplitudes}
When all gravitons $h_1,\dots, h_s$ and gluon traces $\pmb{2},\dots,\pmb{m}$ are collected into the set $\pmb{\mathcal{H}}\equiv\{h_1,\dots,h_s,\pmb{2},\dots,\pmb{m}\}$, a tree level multi-trace EYM amplitude  $A(1,2,\ldots, r|\pmb{2}|\ldots|\pmb{m}\Vert\mathsf{H})$ where $\mathsf{H}\equiv\{h_1,h_2,\dots, h_s\}$ can be briefly expressed by $A(1,2,\ldots, r\Vert\pmb{\mathcal{H}})$.
As proven in \cite{Du:2017gnh}, this multi-trace amplitude satisfies the following recursive expansion relation:
\begin{align}
&A(1,2,\ldots, r\,\Vert\,\pmb{\mathcal{H}})=\sum_{\small\substack{\pmb{\mathcal{H}}\setminus\{\mathcal{H}_a\}\\ \to\text{perms}\,\pmb{\mathcal{H}}_A|\pmb{\mathcal{H}}_B}}\widetilde{\Sl_{\mathsf{Tr}}}\,\biggl[\Sl_{\pmb{\sigma}}C(1,\pmb{\sigma},r)A(1,\pmb{\sigma},r\,\Vert\,\pmb{\mathcal{H}}_B)\biggr].\Label{Eq:MultiTrace}
\end{align}
In the expansion (\ref{Eq:MultiTrace}),
we have picked out an arbitrary element $\mathcal{H}_a$ (a graviton or a gluon trace) from $\pmb{\mathcal{H}}$, which is called the \emph{fiducial element}. Apparently, $\mathcal{H}_a$ can be either a graviton or a gluon trace. On the RHS of \eqref{Eq:MultiTrace}:
\begin{itemize}
\item The first summation is taken over (i). all possible splittings of the set $\pmb{\mathcal{H}}\setminus \{\mathcal{H}_a\}$ into two subsets $ \pmb{\mathcal{H}}_A$, $\pmb{\mathcal{H}}_B$ and (ii). all permutations of elements in  $\pmb{\mathcal{H}}_A$ for a given splitting.

\item For a given splitting of $\pmb{\mathcal{H}}\setminus\{\mathcal{H}_a\}$ and a given permutation of elements in  $\pmb{\mathcal{H}}_A$,
    the summation $\widetilde{\sum}_{\mathsf{Tr}}$ is defined as follows
    \bea
    \text{if $\mathcal{H}_a$ is a graviton:}&&\widetilde{\sum_{\mathsf{Tr}}}\to \sum_{\scriptsize\substack{\{a_i,b_i\}\subset\pmb{t}_i\\\text{for all $\pmb{t}_i\in\pmb{\mathcal{H}}_A$}}}(-1)^{|\pmb{t}_i,a_i,b_i|}\Sl_{\pmb{\beta}_i}\,\Label{Eq:TRTilde1}\\
    \text{if $\mathcal{H}_a=\pmb{t}_0$ is a gluon trace:}&&\widetilde{\sum_{\mathsf{Tr}}}\to \sum_{\scriptsize\substack{a_0\in\pmb{t}_0\\\text{for $a_0\neq b_0$($b_0\in\pmb{t}_0$)}}}(-1)^{|\pmb{t}_0,a_0,b_0|}\Sl_{\pmb{\beta}_0}\sum_{\scriptsize\substack{\{a_i,b_i\}\subset\pmb{t}_i\\\text{for all $\pmb{t}_i\in\pmb{\mathcal{H}}_A$}}}(-1)^{|\pmb{t}_i,a_i,b_i|}\Sl_{\pmb{\beta}_i}. \Label{Eq:TRTilde2}
    \eea
   This means we sum over all possible  choices of the ordered pair of gluons $\{a_i,b_i\}\subset\pmb{t}_i$  for all traces $\pmb{t}_i\in\pmb{\mathcal{H}}_A$. If the fiducial element $\mathcal{H}_a$ is also a trace, namely $\pmb{t}_0$, we should fix an arbitrary gluon $b_0\in \pmb{t}_0$ and then sum over all choices of $a_0\neq b_0$ in this trace. For a given choice of $a_i$ and $b_i$, a gluon trace (including $\mathcal{H}_a$ if it is also a trace) can always be written into the form $a_i,\pmb{X}_i,b_i,\pmb{Y}_i$ where $\pmb{X}_i$ and $\pmb{Y}_i$ are the two ordered sets of gluons separated by $a_i$ and $b_i$. The sign $(-1)^{|\pmb{t}_i,a_i,b_i|}$ in \eqref{Eq:TRTilde1} and \eqref{Eq:TRTilde2} for each trace is defined as $(-1)^{|\pmb{Y}_i|}$ where $|\pmb{Y}_i|$ is the number of elements in $\pmb{Y}_i$. Permutations $\pmb{\beta}_i$ (for a given $\{a_i,b_i\}$) and $\pmb{\beta}_0$ (for a given $a_0\neq b_0$) in \eqref{Eq:TRTilde1} and/or \eqref{Eq:TRTilde2} which satisfy
\bea
\pmb{\beta}_i&\in\mathsf{KK}[\pmb{t}_i,a_i,b_i]&\equiv \pmb{X}_i\shuffle \pmb{Y}_i^T, \Label{Eq:ShuffleInTR}\\
\pmb{\beta}_0&\in\mathsf{KK}[\pmb{t}_0,a_0,b_0]&\equiv\pmb{X}_0\shuffle \pmb{Y}_0^T, \Label{Eq:ShuffleInTR1}
\eea
are also summed over.

\item Supposing that the permutation of elements in $\pmb{\mathcal{H}}_A$ in the first summation (see \eqref{Eq:MultiTrace}) is given by $j_1$, $j_2$,\dots, $j_u$ and the gluon pairs in the second summation are $\{a_i,b_i\}\subset\pmb{t}_i$ ($i$ can be $0$ if the fiducial element $\mathcal{H}_a$ is a gluon trace $\pmb{t}_0$), we sum over all permutations $\pmb{\sigma}$ satisfying
    \bea
    \pmb{\sigma}\in\{2,\dots,r-1\}\shuffle\{j_1,j_2,\dots,j_u,\mathcal{H}_{a}\}.\Label{Eq:sigma}
    \eea
{\emph{Here the traces in the ordered set $\{j_1,j_2,\dots,j_u,\mathcal{H}_{a}\}$ are no longer considered as single elements but considered as proper permutations of all gluons in them. }} Particularly, if  $j_i$ denotes a gluon trace $\pmb{t}_i$ and this trace can be written as $a_i,\pmb{X}_i,b_i,\pmb{Y}_i$ for a given choice of ordered pair $\{a_i,b_i\}\subset\pmb{t}_i$, we should replace  $j_i$ in \eqref{Eq:sigma} by a permutation $\{a_i,\pmb{\beta}_i,b_i\}$ where $\pmb{\beta}_i$ satisfies \eqref{Eq:ShuffleInTR}. Similarly, if the fiducial element $\mathcal{H}_{a}$ is also a trace, say $\pmb{t}_0$, it must be replaced by a permutation $\{a_0,\pmb{\beta}_0,b_0\}$ where $\pmb{\beta}_0$ satisfies \eqref{Eq:ShuffleInTR1}. Then the summation over $\pmb{\sigma}$ in \eqref{Eq:MultiTrace} means summing over all possible shuffle permutations in \eqref{Eq:sigma}.

\item The coefficient $C(1,\pmb{\sigma},r)$ in \eqref{Eq:MultiTrace} is defined as
\bea
\mathcal{E}_{\mathcal{H}_a}\cdot\mathbb{F}_{j_u}\cdot \ldots\cdot\mathbb{F}_{j_i}\cdot Y_{j_1}(\pmb{\sigma}),
\eea
where $\mathcal{E}^{\mu}$ and $\mathbb{F}^{\mu\,\nu}$ are defined by \eqref{Eq:EFY1} and \eqref{Eq:EFY2}.
\bea
Y^{\mu}_{j_1}(\pmb{\sigma})&=&\Biggl\{
                                                                                                                                \begin{array}{cc}
                                                                                                                                 \Sl_{\tiny\sigma^{-1}(l)< \sigma^{-1}(h_x)}k^{\mu}_{l}  &~~~~~~~~\,(\text{if $j_1$ is a graviton $h_x$})  \\
                                                                                                                                 \Sl_{\tiny\sigma^{-1}(l)< \sigma^{-1}(b_i)}k^{\mu}_{l} & ~~~~~~~~~~\,(\text{if $j_1$ is a gluon trace $\pmb{t}_i$})
                                                                                                                                \end{array}.
\eea
\end{itemize}

\subsection{Identities induced from multi-trace EYM amplitudes}\label{Sec:InducedEYMidentities}
Symmetries of multi-trace EYM amplitudes, together with the recursive expansion (\ref{Eq:MultiTrace}), induce nontrivial identities for EYM amplitudes with fewer gravitons and/or gluon traces \cite{Du:2017gnh}. There are two symmetries under consideration in this paper: the gauge invariance condition for a graviton and the cyclic symmetry of a gluon trace.

Muti-trace EYM amplitude satisfies gauge invariance condition, which states that the amplitude has to vanish once half polarization $\epsilon^{\mu}_{h_x}$ of a graviton $h_x$ is replaced by the momentum $k^{\mu}_{h_x}$. If this replacement is performed on the RHS of the recursive expansion  (\ref{Eq:MultiTrace}), we should consider two distinct situations:
\begin{itemize}
\item If the fiducial element $\mathcal{H}_a$ is a graviton $h_x$ (i.e. \eqref{Eq:MultiTrace} is type-I expansion), the gauge invariance condition for $h_x$ induces a nontrivial relation between EYM amplitudes with fewer gravitons
\bea\sum_{\small\substack{\pmb{\mathcal{H}}\setminus h_x\\ \to\text{perms}\,\pmb{\mathcal{H}}_A|\pmb{\mathcal{H}}_B}}\widetilde{\Sl_{\mathsf{Tr}}}\bigg[\Sl_{\pmb{\sigma}}C(1,\pmb{\sigma},r)\Big|_{\epsilon_{h_x}\to k_{h_x}}A(1,\pmb{\sigma},r\,\Vert\,\pmb{\mathcal{H}}_B)\bigg]=0.\Label{Eq:Type-IID}
\eea

\item If $h_x$ is not the fiducial one, it may belong to either $\pmb{\mathcal{H}}_A$ or $\pmb{\mathcal{H}}_B$ in \eqref{Eq:MultiTrace} . Terms with $h_x\in \pmb{\mathcal{H}}_A$ have to vanish due to the antisymmetry of the strength tensor $F_{h_x}^{\mu\nu}\equiv k_{h_x}^{\mu}\epsilon_{h_x}^{\nu}-k_{h_x}^{\nu}\epsilon_{h_x}^{\mu}$, while terms with $h_x\in \pmb{\mathcal{H}}_B$ have to vanish due to gauge invariance of amplitudes with fewer gravitons.
\end{itemize}
Therefore, the only nontrivial identity induced by the gauge invariance condition of a graviton is \eqref{Eq:Type-IID} which is called type-I identity in \cite{Du:2017gnh}.

Another identity (called type-II identity in \cite{Du:2017gnh}) is induced from the expansion (\ref{Eq:MultiTrace}) where the fiducial element $\mathcal{H}_a$ is a gluon trace $\pmb{t}_0$. In particular, we notice that the end element  $b_0\in \pmb{t}_0$ can be chosen arbitrarily in the fiducial trace $\pmb{t}_0$. This arbitrariness is essentially caused by the cyclic symmetry of the trace $\pmb{t}_0$ \cite{Du:2017gnh} and indicates the following identity
\bea
\sum_{\small\substack{\pmb{\mathcal{H}}\setminus \mathcal{H}_a\\ \to\text{perms}\,\pmb{\mathcal{H}}_A|\pmb{\mathcal{H}}_B}}\widetilde{\Sl_{\mathsf{Tr}}}\bigg[\Sl_{\pmb{\sigma}'}C(1,\pmb{\sigma}',r)A(1,\pmb{\sigma}',r\,\Vert\,\pmb{\mathcal{H}}_B)\bigg]=0, \Label{Eq:Type-IIID}
\eea
where $\pmb{\sigma}'$ is defined by
\bea
\pmb{\sigma}'\in\{2,\dots,r-1\}\shuffle\{j_1,j_2,\dots,j_u,a_0,\pmb{\beta}_0\in\mathsf{KK}[\pmb{t}_0,a_0,b_0],\bcancel{b_0}\}.
\eea
Here we removed $b_0$ from the trace $\pmb{t}_0$ first, then shuffled these permutations according to \eqref{Eq:sigma}. The $j_1$, $j_2$, ..., $j_u$ is a permutation of elements in $\pmb{\mathcal{H}}_A$.

When the recursive expansion (\ref{Eq:MultiTrace}) for amplitudes with fewer gravitons and/or gluon traces are applied repeatedly, the two types of relations (\ref{Eq:Type-IID}) and (\ref{Eq:Type-IIID}) respectively turn into the induced identities (\ref{Eq:InducedID1}) and  (\ref{Eq:InducedID2}) for pure YM amplitudes.

\subsection{BCJ relations}

Tree level color-ordered YM amplitudes satisfy the following \emph{traditional BCJ relation} \cite{BjerrumBohr:2009rd,Chen:2011jxa}:
\bea
\Sl_{\pmb{\sigma}\in\pmb{\beta}\,\shuffle\,\pmb{\alpha}}\,\Sl_{l\in\pmb{\beta}}\left(k_{l}\cdot X_l(\pmb{\sigma})\right)A(1,\pmb{\sigma},r)=0,\Label{Eq:BCJRelation}
\eea
where $\pmb{\beta}$ and $\pmb{\alpha}$ are two ordered sets of external gluons, $X_l(\pmb{\sigma})$ denotes the sum of all  momenta of gluons $a\in\{1,\pmb{\sigma}\}\cup\pmb{\beta}$ satisfying $\sigma^{-1}(a)<\sigma^{-1}(l)$.

In \cite{Hou:2018bwm}, the following \emph{graph-based BCJ relation} for YM amplitudes  was proposed
\bea
\Sl_{a\in \mathcal{T}}f^{a}\Sl_{\pmb{\zeta}\in{\mathcal{T}}|_a}\Sl_{\pmb{\sigma}\in\pmb{\zeta}\shuffle\,\pmb{\gamma}}\left[k_{a}\cdot Y_{a}(\pmb{\sigma})\right]\,A\left(1,\pmb{\sigma},r\right)=0.~~\Label{Eq:Graph-based-BCJ}
\eea
Here, $\pmb{\gamma}$ is an arbitrary permutation of elements in $\{2,\dots,r-1\}$ and $\mathcal{T}$ is an arbitrary connected tree graph.
When a node $a$ is chosen as the leftmost element, the tree graph $\mathcal{T}$  establishes permutations $\pmb{\zeta}\in\mathcal{T}|_a$ as follows (i). For two adjacent nodes $x$ and $y$, if $x$ is nearer to $a$ than $y$, we have $\zeta^{-1}(x)<\zeta^{-1}(y)$, (ii). If there are subtree structures attached to a same node, we should shuffle the permutations established by these subtrees together.
The factor $f^a$ is a relative sign depending on the node $a$. This factor is determined by the following steps. {(i).} Choose an arbitrary node $c$ and require $f^c=1$. (ii). For arbitrary two adjacent nodes $c_1$ and $c_2$, we have $f^{c_1}=-f^{c_2}$. As already proven in \cite{Hou:2018bwm}, the graph-based BCJ relation (\ref{Eq:Graph-based-BCJ}) can always be written as a combination of the traditional ones (\ref{Eq:BCJRelation}).

\section{Proof of \eqref{Eq:SplittingTraces}}\label{sec:Appendix}
%
 To prove \eqref{Eq:SplittingTraces}, we focus on a term with a given choice of $\{a_i,b_i\}\subset\pmb{i}$ in \eqref{Eq:SplittingTraces} and prove the following stronger relation:
\bea
	\boxed{\centering
	\includegraphics[width=0.95\textwidth]{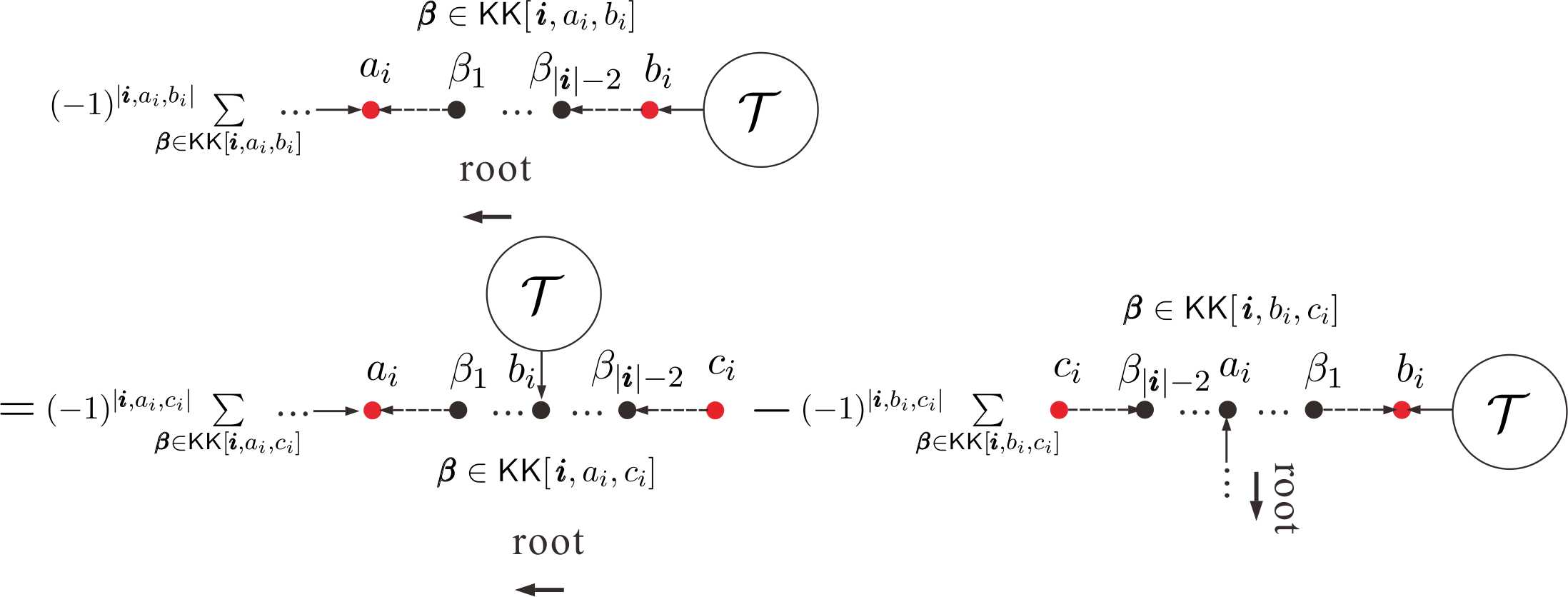}},\nn\Label{Eq:SplittingTraces2}
\eea
where the $\mathcal{T}$ denotes a tree structure which is attached to the node $b_i$. In fact, all these nodes in the trace can be attached by arbitrary tree structures and the relation (\ref{Eq:SplittingTraces2}) still holds. Once all  $\{a_i,b_i\}\subset\pmb{i}$ are summed over, we arrive the relation (\ref{Eq:SplittingTraces}).

Suppose the gluons in trace $\pmb{i}$ are in the cyclic order $d_1,d_2,\dots,d_{|\pmb{i}|}$. Without loss of generality, $a_i$ and $b_i$ are respectively chosen as $a_i=d_j$, $b_i=d_{|\pmb{i}|}$. Then the permutations $\{a_i,\pmb{\beta}\in \mathsf{KK}[\,\pmb{i},a_i,b_i],b_i\}$ on the LHS of \eqref{Eq:SplittingTraces2} are explicitly displayed by
\bea
&&\big\{a_i=d_j,\{d_{j+1},\dots,d_{|\pmb{i}|-1}\}\shuffle\{d_{j-1},\dots,d_{1}\},b_i=d_{|\pmb{i}|},\mathcal{T}|_x\big\},\Label{Eq:APPLHS}\nn
&=&\Bigl\{\big\{a_i=d_j,\{d_{j+1},\dots,d_{|\pmb{i}|-1}\}\shuffle\{d_{j-1},\dots,d_{1}\},b_i=d_{|\pmb{i}|}\big\}\shuffle\mathcal{T}|_x\Bigr\}|_{b_i\prec x}.
\eea
with a sign $(-1)^{|\pmb{i},a_i,b_i|}=(-1)^{j-1}$ (the overall sign coming from $\mathcal{T}|_x$ has been neglected). Here $x$ denotes the nearest to $b_i$ node in $\mathcal{T}$ and $\mathcal{T}|_x$ are the permutations established by the tree $\mathcal{T}$. The notation $b_i\prec x$ means the position of $b_i$ in the permutation is less than that of $x$. The second line of \eqref{Eq:APPLHS}, where  we shuffled $\mathcal{T}|_x$  with the full trace and required the node $b_i$ is always on the left of the node $x$, is apparently equivalent to the first line.

On the RHS of  \eqref{Eq:SplittingTraces2}, we assume that the $c_i$ is chosen as $d_l$ ($l>j$). The case with $l<j$ follows from a similar discussion. Then permutations in the first term on the RHS of \eqref{Eq:SplittingTraces2}  is given by
\bea
\Bigl\{\big\{a_i=d_j,\{d_{j+1},\dots,d_{l-1}\}\shuffle\{d_{j-1},\dots,d_1,b_i=d_{|\pmb{i}|},\dots,d_{l+1}\},c_i=d_l\bigr\}\shuffle\mathcal{T}|_x\Bigr\}|_{b_i\prec x},\Label{Eq:APPRHS1}
\eea
with the sign $(-1)^{|\pmb{i},a_i,c_i|}=(-1)^{|\pmb{i}|-l+j-1}$. The permutations in the second term of the RHS of \eqref{Eq:SplittingTraces} are given by the following two steps:
\begin{itemize}
\item First shuffle the gluons in the trace $\pmb{i}$ so that $b_i=d_{|\pmb{i}|}$ and $c_i=d_l$ become the two ends of the trace. Then the permutations $\pmb{\beta}\in \mathsf{KK}[\,\pmb{i},b_i,c_i]$ are explicitly given by
    \bea
    \pmb{\beta}\in\{d_{|\pmb{i}|-1},\dots,d_{l+1}\}\shuffle\{d_1,\dots,d_{l-1}\}.\Label{Eq:AppBeta}
    \eea
    which is associated with a sign $(-1)^{|\pmb{i},b_i,c_i|}=(-1)^{|\pmb{i}|-l-1}$. These permutations can be classified according to the relative orders between the $a_i=d_j$ $(d_j\in \{d_1,\dots,d_{l-1}\})$ and elements in the set $\{d_{|\pmb{i}|-1},\dots,d_{l+1}\}$:
    \bea
    \pmb{\beta}^{{(1)}}&\in&\bigl\{d_1,\dots,d_{j-1},d_j,\{d_{j+1},\dots,d_{l-1}\}\shuffle\{d_{|\pmb{i}|-1},\dots,d_{l+1}\}\bigr\}\nn
     \pmb{\beta}^{{(2)}}&\in&\bigl\{\{d_1,\dots,d_{j-1}\}\shuffle\{d_{|\pmb{i}|-1}\},d_j,\{d_{j+1},\dots,d_{l-1}\}\shuffle\{d_{|\pmb{i}|-2},\dots,d_{l+1}\}\bigr\}\nn
     &\dots&\nn
    \pmb{\beta}^{{(q)}}&\in&\bigl\{\{d_1,\dots,d_{j-1}\}\shuffle\{d_{|\pmb{i}|-1},\dots,d_{|\pmb{i}|-q+1}\},d_j,\{d_{j+1},\dots,d_{l-1}\}\shuffle\{d_{|\pmb{i}|-q},\dots,d_{l+1}\}\bigr\}\nn
     &\dots&\nn
     \pmb{\beta}^{(|\pmb{i}|-l-1)}&\in&\bigl\{\{d_1,\dots,d_{j-1}\}\shuffle\{d_{|\pmb{i}|-1},\dots,d_{l+2}\},d_j,\{d_{j+1},\dots,d_{l-1}\}\shuffle\{d_{l+1}\}\bigr\}\nn
     \pmb{\beta}^{(|\pmb{i}|-l)}&\in&\bigl\{\{d_1,\dots,d_{j-1}\}\shuffle\{d_{|\pmb{i}|-1},\dots,d_{l+1}\},d_j,d_{j+1},\dots,d_{l-1}\bigr\}.\Label{Eq:AppBeta1}
    \eea

\item For any given permutation $\pmb{\beta}^{(q)}$ in \eqref{Eq:AppBeta1}, the permutations $\pmb{\sigma}^{(q)}$ established by the second term on the RHS of \eqref{Eq:SplittingTraces2} are obtained by shuffling the two branches attached to the node $a_i=d_i$ together:
\bea
\pmb{\sigma}^{(q)}&\in&\biggl\{\Bigl\{a_i=d_j,\bigl\{\{d_{j+1},\dots,d_{l-1}\}\shuffle\{d_{|\pmb{i}|-q},\dots,d_{l+1}\},c_i=d_l\bigr\}\nn
&&~~~~~~~~~~~~~~~~~~~~~~\shuffle\bigl\{\{d_{j-1},\dots,d_{1}\}\shuffle\{d_{|\pmb{i}|-q+1},\dots,d_{|\pmb{i}|-1}\},b_i=d_{|\pmb{i}|}\bigr\}\Bigr\}\shuffle\mathcal{T}|_x\biggr\}\Big|_{b_i\prec x}\Label{Eq:AppSigma}
\eea
with a sign $(-1)^{|\pmb{i}|-l-1}(-1)^{\mathcal{N}^{(q)}_{a_i}}$. Here $\mathcal{N}^{(q)}_{a_i}$ denotes the number of arrows pointing away from root in the trace $\pmb{i}$.
\end{itemize}
According to the relative orders between $d_{|\pmb{i}|-q}$ and $d_{|\pmb{i}|-q+1}$, permutations
$\pmb{\sigma}^{(q)}$ in \eqref{Eq:AppSigma} splits into $\pmb{\sigma}^{(q)}_{\text{A}}\equiv\pmb{\sigma}^{(q)}|_{d_{|\pmb{i}|-q}\prec d_{|\pmb{i}|-q+1}}$ and $\pmb{\sigma}^{(q)}_{\text{B}}\equiv\pmb{\sigma}^{(q)}|_{d_{|\pmb{i}|-q}\succ d_{|\pmb{i}|-q+1}}$. It is easy to see
\bea
\pmb{\sigma}^{(q)}_{\text{A}}&=&\pmb{\sigma}^{(q+1)}_{\text{B}},~~~~\mathcal{N}^{(q)}_{a_i}=\mathcal{N}^{(q+1)}_{a_i}+1.
\eea
Hence the contributions from $\pmb{\sigma}^{(q)}_{\text{A}}$ and $\pmb{\sigma}^{(q+1)}_{\text{B}}$ for all $q=1,\dots,|\pmb{i}|-l-1$ must cancel one another. The remaining nonzero terms are the two boundaries
\bea
\pmb{\sigma}^{(1)}_{\text{B}}&\in&\biggl\{\Bigl\{a_i=d_j,\bigl\{\{d_{j+1},\dots,d_{l-1}\}\shuffle\{d_{|\pmb{i}|-1},\dots,d_{l+1}\},c_i=d_l\bigr\}\nn
&&~~~~~~~~~~~~~~~~~~~~~~~~~~~~~~~~~~~~~~~~~~~~~~~~~~~~~~~~~~~\shuffle\bigl\{d_{j-1},\dots,d_{1},b=d_{|\pmb{i}|}\bigr\}\Bigr\}\big|_{d_{|\pmb{i}|-1}\succ d_{|\pmb{i}|}}\shuffle \mathcal{T}|_x\biggr\}\Big|_{b_i\prec x}\nn
&=&\Bigl\{\bigl\{a_i=d_j,\{d_{j+1},\dots,d_{l-1}\}\shuffle\{d_{j-1},\dots,d_{1},b_i=d_{|\pmb{i}|},d_{|\pmb{i}|-1},\dots,d_{l+1}\},c_i=d_l\bigr\}\shuffle\mathcal{T}|_x\Bigr\}\big|_{b_i\prec x}.\nn
\pmb{\sigma}^{(|\pmb{i}|-l)}_{\text{A}}&\in&\biggl\{\Bigl\{a_i=d_j,\{d_{j+1},\dots,d_{l-1},c_i=d_l\bigr\}\nn
&&~~~~~~~~~~~~~~~~~~~~~~~~~~~~~~~~\shuffle\bigl\{\{d_{j-1},\dots,d_{1}\}\shuffle\{d_{l+1},\dots,d_{|\pmb{i}|-1}\},b_i=d_{|\pmb{i}|}\bigr\}\Bigr\}\big|_{d_l\prec d_{l+1}}\shuffle\mathcal{T}|_x\biggr\}\Big|_{b_i\prec x}\nn
&=&\Bigl\{\bigl\{a_i=d_j,\{d_{j-1},\dots,d_{1}\}\shuffle\{d_{j+1},\dots,d_{|\pmb{i}|-1}\},b_i=d_{|\pmb{i}|}\bigr\}\shuffle\mathcal{T}|_x\Bigr\}\big|_{b_i\prec x},\Label{Eq:AppSigma1}
\eea
with the signs  $(-1)^{|\pmb{i}|-l-1}(-1)^{\mathcal{N}^{(1)}_{a_i}}=(-1)^{|\pmb{i}|-l-1+j}$ and $(-1)^{|\pmb{i}|-l-1}(-1)^{\mathcal{N}^{(|\pmb{i}|-l)}_{a_i}}=(-1)^{j}$ respectively. An extra minus which is introduced from the second term of \eqref{Eq:SplittingTraces2} must also be taken into account. Hence the permutations $\pmb{\sigma}^{(1)}_{\text{B}}$ are same with the permutations in \eqref{Eq:APPRHS1}, with an opposite sign. As a result, they cancel with each other. The remaining permutations $\pmb{\sigma}^{(|\pmb{i}|-l)}_{\text{A}}$ are nothing but the permutations (\ref{Eq:APPLHS}) with the corrected sign.
Therefore, we have proven the relation (\ref{Eq:SplittingTraces2}) for a given $\{a_i,b_i\}$. After summing over all possible choices of the $\{a_i,b_i\}$ pairs, the proof of \eqref{Eq:SplittingTraces} is completed.

\begin{figure}
	\centering
	\includegraphics[width=0.97\textwidth]{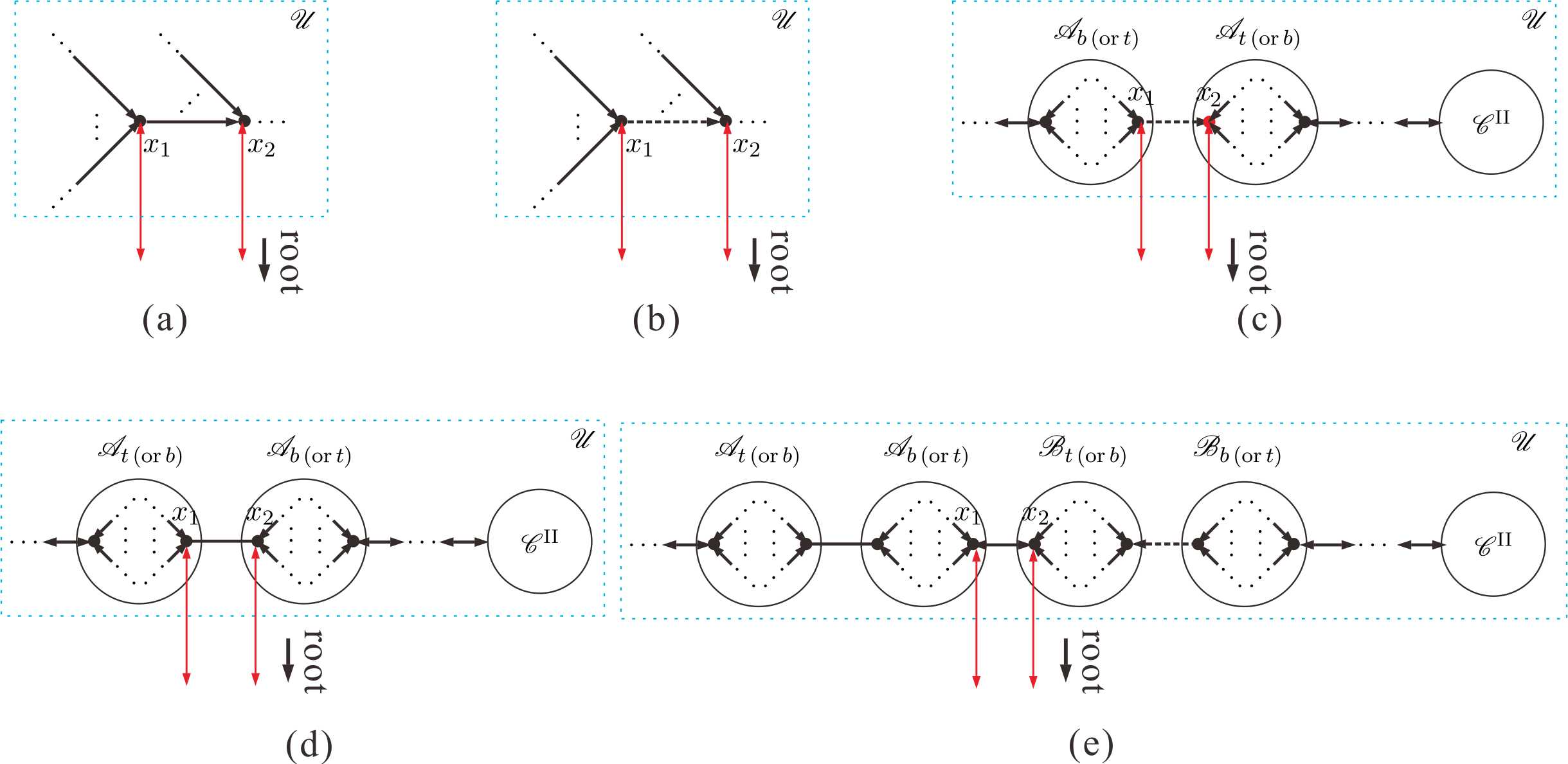}
	\caption{All possible situations when choosing $x\in \mathscr{U}$ as two adjacent nodes $x_1$ and $x_2$.}\label{Fig:SignMulti}
\end{figure}

{\bf\emph{Comments on the proof:}} In the above proof, the tree structure $\mathcal{T}|_x$ is always treated separately from the trace in each step (i.e. it is shuffled with the full trace with a proper constraint, as shown in eqs. (\ref{Eq:APPLHS}), (\ref{Eq:APPRHS1}), (\ref{Eq:AppSigma}), (\ref{Eq:AppSigma1})). Thus manipulations on nodes inside the trace are independent of $\mathcal{T}|_x$. This observation allows us to generalize \eqref{Eq:SplittingTraces} to cases where more tree structures are attached to nodes in the trace $\pmb{i}$  straightforwardly.

\section{The sign in \eqref{Eq:InducedIDUnified4}}\label{sec:Sign}

In \eqref{Eq:InducedIDUnified4}, the graph $\mathcal{G}$ is constructed by connecting $x\in\mathscr{U}$ and $y\in\mathscr{L}$ via a type-3 line. As mentioned before, the sign for such a graph is dependent of the choice of $x$ but independent of the choice of $y$. Now we show that
two graphs with choosing adjacent $x\in\mathscr{U}$ must have opposite signs. To see this, we study all possible structures presented by \figref{Fig:SignMulti} (a)-(e), where $x=x_1$ is adjacent to $x=x_2$ ($x_1,x_2\in\mathscr{U}$), as follows.

\noindent~\textbf{(i).} As shown by \figref{Fig:SignMulti} (a), nodes $x_1$ and $x_2$ connected by a type-3 line  must belong to a same (top or bottom) side of a type-IA or type-IB component. In this case, each one of $S(\mathscr{U}_x)$, $Tr(\mathscr{U}_x,\mathscr{L})$ and $\mathcal{N}(\mathscr{L})$ is the same for choosing two adjacent $x$'s. But the numbers $\mathcal{N}(\mathscr{U}_x)$ in \eqref{Eq:Sign} for $x=x_1$ and $x=x_2$ differ by one because the type-2 line between $x_1$ and $x_2$ has opposite directions for these two cases.

\noindent~\textbf{(ii).} If $x_1$ and $x_2$ are connected by a type-4 line, they must belong to a same type-IB component.  If the type-4 line is not the kernel of the type-IB component, the structure is given by \figref{Fig:SignMulti} (b). In this case, $x_1$ and $x_2$ must belong to a same side of the component. The numbers $\mathcal{N}(\mathscr{U}_x)$ for choosing $x=x_1$ and $x=x_2$ should differ by one because the type-4 line between $x_1$ and $x_2$ has opposite directions. Each of  $S(\mathscr{U}_x)$, $Tr(\mathscr{U}_x,\mathscr{L})$ and $\mathcal{N}(\mathscr{L})$ in \eqref{Eq:Sign} is the same for both choices of $x$.

\noindent~\textbf{(iii).} If the two adjacent nodes $x_1$ and $x_2$ connected by a type-4 line belong to opposite sides of a type-IB component,  the type-4 line must be the kernel of the type-IB component (as shown by \figref{Fig:SignMulti} (c)). In this case, each of $\mathcal{N}(\mathscr{U}_x)$, $Tr(\mathscr{U}_x,\mathscr{L})$ and $S(\mathscr{U}_x)$ for $x=x_1$ and $x=x_2$ differ by one but $\mathcal{N}(\mathscr{L})$ is the same.

\noindent~\textbf{(iv).} If $x_1$ and $x_2$ are connected by a type-1 line which must be the kernel of a type-IA component (as shown by \figref{Fig:SignMulti} (d)), only the number $S(\mathscr{U}_x)$ for choosing $x=x_1$ and $x=x_2$ differ by one. Any of $Tr(\mathscr{U}_x,\mathscr{L})$, $\mathcal{N}(\mathscr{U}_x)$ and $\mathcal{N}(\mathscr{L})$ is the same for the two choices of $x$.

\noindent~\textbf{(v).} If $x_1$ and $x_2$ are connected by a type-3 line as shown by \figref{Fig:SignMulti} (e), only  $S(\mathscr{U}_x)$ for  choosing $x=x_1$ and $x=x_2$ differ by one. Any other number is the same for the two choices.

Therefore, in all the above cases, $S(\mathscr{U}_{x})+Tr(\mathscr{U}_{x},\mathscr{L})+\mathcal{N}(\mathscr{U}_{x})+\mathcal{N}(\mathscr{L})+1$ for choosing $x=x_1$ and $x=x_2$, where $x_1,x_2\in\mathscr{U}$ are two adjacent nodes, must differ by an odd number. In other words, two graphs with adjacent $x$'s have opposite signs.

 \bibliographystyle{JHEP}
\bibliography{NoteONGaugeIdentity}

\end{document}